# Title page

Names of the authors: F. Tárkányi[1], A. Hermanne[2], F. Ditrói[1], S. Takács[1]

Title: **ACTIVATION CROSS SECTIONS OF PROTON INDUCED NUCLEAR REACTIONS ON NEODYMIUM UP TO 65 MEV**

Affiliation(s) and address(es) of the author(s):

[1] Institute for Nuclear Research, Hungarian Academy of Sciences (ATOMKI), Debrecen, Hungary
[2] Cyclotron Laboratory, Vrije Universiteit Brussel (VUB), Brussels, Belgium

E-mail address of the corresponding author: ditroi@atomki.hu

# ACTIVATION CROSS SECTIONS OF PROTON INDUCED NUCLEAR REACTIONS ON NEODYMIUM UP TO 65 MEV

**F. Tárkányi[1], A. Hermanne[2] . F. Ditrói[1*], S. Takács[1]**

[1] *Institute for Nuclear Research, Hungarian Academy of Sciences (ATOMKI), Debrecen, Hungary*

[2] *Cyclotron Laboratory, Vrije Universiteit Brussel (VUB), Brussels, Belgium*

## ABSTRACT

In the frame of a systematic study of the activation cross sections of charged particle induced nuclear reactions on rare earths for production of therapeutic radionuclides, proton induced reactions on neodymium were measured up to 65 MeV energy, above 45 MeV for the first time. The excitation functions of the $^{nat}Nd(p,x)^{150,149,148m,148g,146,144,143,141}Pm$, $^{149,147,141,140,139m,138}Nd$, $^{142,138m}Pr$ and $^{139g}Ce$ nuclear reactions were assessed by using stacked foil activation technique and high resolution γ-spectrometry. The excitation functions were compared to the theoretical predictions, available in the TENDL-2015 library based on latest version of the TALYS code. The application of the data for medical isotope production is shortly discussed.

[*] Corresponding author: ditroi@atomki.hu

## 1. INTRODUCTION

This work was performed in the frame of a systematic study of charged particle induced production routes of medical radioisotopes. Numerous radioisotopes of lanthanides exist in standard medical practice or are considered as emerging. Among new nuclides suitable for therapeutic purposes [1-9] the radionuclide $^{149}$Pm ($T_{1/2}$ = 53.1 h) - $^{149}$Nd ($T_{1/2}$ = 1.728 h) generator pair, $^{140}$Nd ($T_{1/2}$ = 3.37 d), and $^{139}$Ce($T_{1/2}$ =137.641 d) were found to offer unique properties suitable for therapy and the daughter nuclide $^{140}$Pr ($T_{1/2}$ = 3.4 min) offers the additional advantage of in vivo localization via positron emission tomography (PET). We already investigated the production routes of these radionuclides by using deuteron induced reactions on neodymium [10]. We also have investigated possible production routes of these radionuclides via proton and deuteron induced reactions on La [11], Ce [12,13] and Pr [14,15].

Searching the literature for activation cross sections of proton induced nuclear reactions on neodymium, three works have been found up to 38 MeV, i.e. by Lebeda et al. [16,17], and the recent data of the Korean group up to 45 MeV protons [18-20].

Thick target yield data up to 12 MeV were reported by Muminov et al. [21] for production of $^{142}$Pm and by Dmitriev and Molin at 22 MeV for production of $^{142,143,144,148}$Pm [22].

## 2. EXPERIMENT AND DATA EVALUATION

For measurements the well-established activation method, stacked foil irradiation technique and high resolution γ-spectrometry were used. Neodymium metal foil targets, interleaved with Al and Ti beam monitor foils and Hg containing targets throughout the stack, were irradiated at UCL (LLN) cyclotron at 65 MeV and at VUB cyclotron at 15 MeV proton beam. (see Table 1). Complete excitation functions were measured for the monitor reactions to control the beam intensity and the energy. The main experimental parameters and methods of data evaluation are collected in Table 1 and Table 2. The comparison of the re-measured monitor reactions and the recommended data were reported in our recent paper on activation cross sections of proton induced nuclear reactions on mercury [23]. The decay characteristic of the investigated reaction products and the contributing reactions are summarized in Table 3. The irradiations were done in 2015. The original irradiation plan contains a 36 MeV irradiation, but due to target problems the irradiation was not done finally, which resulted 14-29 MeV missing energy range.

**Table 1.** Main experimental parameters

| Incident particle | Proton | Proton |
|---|---|---|
| Method | Stacked foil | Stacked foil |
| Series | ser. 1 | ser. 2 |
| Target composition (foils) and thickness | Al 9.15 μm<br>HgS 37.05-99.14 $mg/cm^2$<br>Al 50 μm<br>Nd 29.8 μm<br>Ti 9.15 μm<br>(repeated 11 times) | Al 9.15 μm<br>HgS 107.52-197.91 $mg/cm^2$<br>Al 377 μm<br>Nd 29.8 μm<br>Al 377 μm<br>(repeated 24 times) |
| Number of Nd targets | 10 | 24 |
| Accelerator | CGR 560 cyclotron, Vrije Universiteit Brussel (VUB), Brussels, Belgium | Cyclone 110 cyclotron Université Catholique Louvain la Neuve (LLN), Belgium |
| Primary energy | 15 MeV | 65 MeV |
| Covered energy range | 14.32 - 0 MeV | 64 - 28.7 MeV |
| Irradiation time | 60 min | 60 min |
| Beam current | 132 nA | 50 nA |
| Monitor reaction, [recommended values] | $^{nat}Ti(p,x)^{48}V$ reaction [24] (re-measured over the whole energy range) | $^{27}Al(p,x)^{24}Na$ reaction [24] (re-measured over the whole energy range) |
| Monitor, thickness | $^{nat}Ti$, 9.15 μm | $^{nat}Al$, 370 μm |
| detector | HPGe | HPGe |
| Chemical separation | no | no |
| γ-spectra measurements | 4 series | 4 series |
| Typical cooling times | 1.2-4.8 (5 cm)<br>46.63-52.6 h (5 cm) | 7.3-25.9 h (15 cm)<br>124.8-165.3 h (5 cm) |
| (target-detector distances) | 239.3-266.8 h (5 cm)<br>2282.4-2572.0 h (5 cm) | 431.8-579.4 h (5 cm)<br>2737.0-3486.9 (5 cm) |

**Table 2.** Main parameters and methods of the data evaluation (with references)

| | | |
|---|---|---|
| Gamma spectra evaluation | Genie 2000, Forgamma | [25,26] |
| Determination of beam intensity | Faraday cup (preliminary)<br>Fitted monitor reaction (final) | <br>[27] |
| Decay data (see Table 2) | NUDAT 2.6 | [28] |
| Reaction Q-values(see Table 3) | Q-value calculator | [29] |
| Determination of beam energy | Andersen (preliminary)<br>Fitted monitor reaction (final) | [30]<br>[22] |
| Uncertainty of energy | Cumulative effects of possible uncertainties (primary energy, target thickness, energy straggling, correction to monitor reaction) | |
| Cross sections | elemental cross sections | |
| Uncertainty of cross sections | Sum in quadrature of all individual contributions:<br>(beam current (7%),<br>beam-loss corrections (max. 1.5%),<br>target thickness (1%),<br>detector efficiency (5%),<br>photo peak area determination<br>counting statistics (1-20 %) | [31] |
| Yield | Physical yield | [32,33] |

**Table 3.** Decay characteristic of the investigated reaction products and the contributing reactions

| Nuclide<br>(E-level (keV))<br>Decay path | Half-life | $E_\gamma$ (keV) | $I_\gamma$ (%) | Contributing reactions | Q-value (keV)<br>GS-GS |
|---|---|---|---|---|---|
| **$^{150}$Pm**<br>$\beta^-$:100% | 2.698 h | 333.92<br>406.51<br>831.85<br>876.41<br>1165.77<br>1324.51 | 68<br>5.6<br>11.9<br>7.3<br>15.8<br>17.5 | $^{150}$Nd(p,n) | -865.0 |
| **$^{149}$Pm**<br>$\beta^-$:100% | 53.08 h | 285.95 | 3.1 | $^{150}$Nd(p,2n) | -6469.07 |
| **$^{148m}$Pm**<br>137.93 keV<br>IT: 4.2 %<br>$\beta^-$: 95.8 % | 41.29 d | 432.78<br>550.27<br>629.97<br>725.70<br>915.33<br>1013.81 | 5.35<br>94.9<br>89.0<br>32.8<br>17.17<br>20.30 | $^{148}$Nd(p,n)<br>$^{150}$Nd(p,3n) | -1325.08<br>-13739.0 |
| **$^{148g}$Pm**<br>$\beta^-$: 100 % | 5.368 d | 550.27<br>914.85<br>1465.12 | 22.0<br>11.5<br>22.2 | $^{148}$Nd(p,n)<br>$^{150}$Nd(p,3n) | -1325.08<br>-13739.0 |
| **$^{146}$Pm**<br>ε: 66.0 %<br>$\beta^-$: 34.0 % | 5.53 y | 453.88<br>735.93<br>747.24 | 65.0<br>22.5<br>34.0 | $^{146}$Nd(p,n)<br>$^{148}$Nd(p,3n)<br>$^{150}$Nd(p,5n) | -2253.89<br>-14878.6<br>-27292.51 |
| **$^{144}$Pm**<br>ε: 100 % | 363 d | 476.78<br>618.01<br>696.49 | 43.8<br>98<br>99.49 | $^{144}$Nd(p,n)<br>$^{145}$Nd(p,2n)<br>$^{146}$Nd(p,3n)<br>$^{148}$Nd(p,5n)<br>$^{150}$Nd(p,7n) | -3114.27<br>-8869.57<br>-16434.81<br>-29059.52<br>-41473.42 |
| **$^{143}$Pm**<br>ε: 100 % | 265 d | 741.98 | 38.5 | $^{143}$Nd(p,n)<br>$^{144}$Nd(p,2n)<br>$^{145}$Nd(p,3n)<br>$^{146}$Nd(p,4n)<br>$^{148}$Nd(p,6n)<br>$^{150}$Nd(p,8n) | -1824.01<br>-9641.05<br>-15396.35<br>-22961.59<br>-35586.29<br>-48000.2 |
| **$^{141}$Pm**<br>**ε:** 100 **%** | 20.9 min | 193.67<br>886.22<br>1223.26 | 1.61<br>2.44<br>4.7 | $^{142}$Nd(p,2n)<br>$^{143}$Nd(p,3n)<br>$^{144}$Nd(p,4n)<br>$^{145}$Nd(p,5n)<br>$^{146}$Nd(p,6n)<br>$^{148}$Nd(p,8n) | -14280.6<br>-20404.2<br>-28221.2<br>-33976.5<br>-41541.8<br>-54166.5 |
| **$^{149}$Nd**<br>$\beta^-$: 100 % | 1.728 h | 114.31<br>211.31 | 19.2<br>25.9 | $^{150}$Nd(p,pn)<br>$^{149}$Pr decay | -7375.12<br>-9928.7 |
| | | 270.17<br>326.55<br>423.55<br>540.51<br>654.83 | 10.7<br>4.56<br>7.4<br>6.6<br>8.0 | | |
| **$^{147}$Nd**<br>$\beta^-$: 100 % | 10.98 d | 91.105<br>319.41<br>531.02 | 28.1<br>2.13<br>13.4 | $^{148}$Nd(p,pn)<br>$^{150}$Nd(p,p3n)<br>$^{147}$Pr decay | -7332.5<br>-19746.4<br>-9252.8 |
| **$^{141}$Nd**<br>ε: 100 % | 2.49 h | 145.45<br>1126.91<br>1147.30<br>1292.64 | 0.24<br>0.80<br>0.307<br>0.46 | $^{142}$Nd(p,pn)<br>$^{143}$Nd(p,p2n)<br>$^{144}$Nd(p,p3n)<br>$^{145}$Nd(p,p4n)<br>$^{146}$Nd(p,p5n)<br>$^{148}$Nd(p,p7n)<br>$^{150}$Nd(p,p9n)<br>$^{141}$Pm decay | -9827.8<br>-15951.36<br>-23768.41<br>-29523.71<br>-37088.95<br>-49713.64<br>-62127.55<br>-14280.6 |
| **$^{140}$Nd**<br>**ε:** 100 % | 3.37 d | no γ<br>through<br>$^{140}$Pr | | $^{142}$Nd(p,p2n)<br>$^{143}$Nd(p,p3n)<br>$^{144}$Nd(p,p4n)<br>$^{145}$Nd(p,p5n)<br>$^{146}$Nd(p,p6n)<br>$^{148}$Nd(p,p8n)<br>$^{140}$Pm decay | -17838.7<br>-23962.3<br>-31779.4<br>-37534.7<br>-45099.9<br>-57724.6<br>-24666.3 |
| **$^{139m}$Nd**<br>231.15 keV<br>ε: 88.2 %<br>IT: 11.8 % | 5.50 h | 113.87<br>708.1<br>738.2<br>827.8<br>982.2 | 40<br>26<br>35<br>10.3<br>26 | $^{142}$Nd(p,p3n)<br>$^{143}$Nd(p,p4n)<br>$^{144}$Nd(p,p5n)<br>$^{145}$Nd(p,p6n)<br>$^{146}$Nd(p,p7n)<br>$^{148}$Nd(p,p9n)<br>$^{139}$Pm decay | -28149.2<br>-34272.8<br>-42089.8<br>-47845.1<br>-55410.3<br>-68035.0<br>-44077.0 |
| **$^{138}$Nd**<br>ε: 100 % | 5.04 h | 325.76 | 2.9 | $^{142}$Nd(p,p4n)<br>$^{143}$Nd(p,p5n)<br>$^{144}$Nd(p,p6n)<br>$^{145}$Nd(p,p7n)<br>$^{146}$Nd(p,p8n)<br>$^{138}$Pm decay | -36216.9<br>-42340.4<br>-50157.5<br>-55912.8<br>-63478.0<br>-44077.0 |
| **$^{142}$Pr**<br>ε: 0.0164 %<br>$\beta^-$:99.9836% | 19.12 h | 1575.6 | 3.7 | $^{143}$Nd(p,2p)<br>$^{144}$Nd(p,2pn)<br>$^{145}$Nd(p,2p2n)<br>$^{146}$Nd(p,2p3n)<br>$^{148}$Nd(p,2p5n)<br>$^{150}$Nd(p,2p7n) | -7222.44<br>-15319.89<br>-21075.2<br>-28640.43<br>-41265.13<br>-53679.03 |
| **$^{140}$Pr**<br>ε: 100 % | 3.39 min | 306.9<br>1596.1 | 0.147<br>0.49 | daughter of<br>$^{140}$Nd | |
| **$^{139}$Pr**<br>ε: 100 % | 4.41 h | 255.11<br>1347.33<br>1630.67 | 0.236<br>0.473<br>0.343 | $^{142}$Nd(p,2p2n)<br>$^{143}$Nd(p,2p3n)<br>$^{144}$Nd(p,2p4n)<br>$^{145}$Nd(p,2p5n)<br>$^{146}$Nd(p,2p6n)<br>$^{148}$Nd(p,2p8n)<br>$^{139}$Nd decay | -24560.36<br>-30683.94<br>-38500.97<br>-44256.27<br>-51821.51<br>-64446.2<br>-28149.2 |
| **$^{138m}$Pr**<br>364 keV<br>ε: 100 % | 2.12 h | 302.7<br>390.9<br>547.5<br>788.7<br>1037.8 | 80<br>6.1<br>5.2<br>100<br>101 | $^{142}$Nd(p,2p3n)<br>$^{143}$Nd(p,2p4n)<br>$^{144}$Nd(p,2p5n)<br>$^{145}$Nd(p,2p6n)<br>$^{146}$Nd(p,2p7n) | -34321.1<br>-40444.7<br>-48261.7<br>-54017.0<br>-61582.2 |
| **$^{139}$Ce**<br>ε: 100 % | 137.641 d | 165.8575 | 80 | $^{142}$Nd(p,3pn)<br>$^{143}$Nd(p,3p2n)<br>$^{144}$Nd(p,3p3n)<br>$^{145}$Nd(p,3p4n)<br>$^{146}$Nd(p,3p5n)<br>$^{148}$Nd(p,3p7n)<br>$^{139}$Pr decay | -21648.94<br>-27772.52<br>-35589.55<br>-41344.85<br>-48910.08<br>-61534.77<br>-24560.36 |

When complex particles are emitted instead of individual protons and neutrons the Q-values have to be decreased by the respective binding energies of the compound particles: np-d, +2.2 MeV; 2np-t, +8.48 MeV;

$^{nat}$Nd isotopic abundances: $^{142}$Nd (27.13 %), $^{143}$Nd (12.18 %), $^{144}$Nd (23.80 %),$^{145}$Nd (5.30 %), $^{146}$Nd (17.19 %), $^{148}$Nd (5.76 %), $^{150}$Nd (5.64 %)

## 3. RESULTS

### 3.1 Cross sections

The cross-sections for all the reactions investigated are shown in Figs. 1–17 and the numerical values are collected in Tables 4-5. The experimental data are also compared with the cross section data reported in the TALYS based [34] TENDL-2015 On-line Data Library [35].

In some cases the experimental results of these investigations have larger uncertainties, due to several reasons. The high energy accelerator has limited availability, therefore during the day of the experiment a large number of the target foils were irradiated and the gamma spectra were measured in another institute situated in another city. During the time period of the irradiations and the time period required to the target transport, and for the separation of the targets from the stacks the short-lived isotopes have decayed out. We could use only two detectors simultaneously, therefore a few series of targets should stay on waiting list (or only every second target foil were measured in the first series). The cooling times for the different series of measurements are indicated in Table 1. As a detailed discussion on production routes and contributing reactions can be found in the earlier report of Lebeda et al. [16,17] and Yang et al. [18,20,19], we discuss them only briefly here to avoid repetitions.

**Table 4.** Experimental cross-sections for the $^{nat}Nd(p,x)^{150,149,148m,148g,146,144,143,141}Pm$ reactions

| Bombarding energy | | $^{150}$Pm | | $^{149}$Pm | | $^{148m}$Pm | | $^{148g}$Pm | | $^{146}$Pm | | $^{144}$Pm | | $^{143}$Pm | | $^{141}$Pm | |
|---|---|---|---|---|---|---|---|---|---|---|---|---|---|---|---|---|---|
| E (MeV) | ΔE (MeV) | σ (mb) | Δσ (mb) | σ (mb) | Δσ (mb) | σ (mb) | Δσ (mb) | σ (mb) | Δσ (mb) | σ (mb) | Δσ (mb) | σ (mb) | Δσ (mb) | σ (mb) | Δσ (mb) | σ (mb) | Δσ (mb) |
| **Series 1** | | | | | | | | | | | | | | | | | |
| 14.32 | 0.20 | 1.45 | 0.16 | 44.67 | 5.02 | 0.80 | 0.10 | 1.77 | 0.24 | 7.94 | 1.13 | 75.49 | 8.52 | 181.14 | 20.42 | 1.86 | 0.24 |
| 13.24 | 0.24 | 1.49 | 0.17 | 37.19 | 4.19 | 0.74 | 0.10 | 1.33 | 0.20 | 8.07 | 1.14 | 69.52 | 7.85 | 171.00 | 19.28 | 0.36 | 0.12 |
| 12.03 | 0.29 | 1.77 | 0.20 | 33.40 | 3.76 | 0.91 | 0.11 | 1.99 | 0.27 | 14.23 | 1.77 | 78.41 | 8.84 | 158.28 | 17.84 | | |
| 10.88 | 0.34 | 2.15 | 0.24 | 26.07 | 2.94 | 1.19 | 0.14 | 2.44 | 0.31 | 25.31 | 3.04 | 89.98 | 10.10 | 98.30 | 11.09 | | |
| 9.45 | 0.39 | 3.01 | 0.34 | 13.23 | 1.50 | 1.64 | 0.19 | 4.29 | 0.52 | 42.31 | 4.80 | 70.50 | 7.95 | 29.90 | 3.43 | | |
| 7.80 | 0.46 | 3.58 | 0.40 | 2.15 | 0.25 | 1.24 | 0.15 | 4.77 | 0.59 | 20.28 | 2.44 | 19.25 | 2.17 | 9.07 | 1.12 | | |
| 6.00 | 0.53 | 0.74 | 0.08 | | | 0.14 | 0.02 | 0.67 | 0.09 | 2.03 | 0.34 | 2.26 | 0.29 | 1.72 | 0.30 | | |
| **Series 2** | | | | | | | | | | | | | | | | | |
| 64.04 | 0.20 | 0.38 | 0.18 | 9.28 | 2.09 | 1.95 | 0.22 | | | 15.10 | 2.01 | 45.13 | 6.27 | 70.85 | 11.67 | | |
| 62.22 | 0.26 | | | 8.32 | 4.26 | 2.18 | 0.26 | | | 16.81 | 2.12 | 52.48 | 7.49 | 52.17 | 10.55 | | |
| 60.36 | 0.32 | 0.26 | 0.08 | 8.60 | 3.86 | 2.18 | 0.27 | 1.96 | 0.88 | 16.53 | 2.02 | 44.23 | 6.66 | 68.97 | 12.47 | | |
| 58.44 | 0.38 | | | 7.53 | 4.28 | 2.41 | 0.28 | | | 18.45 | 2.30 | 50.82 | 7.45 | 77.21 | 12.99 | | |
| 56.47 | 0.45 | 0.22 | 0.06 | 6.35 | 2.93 | 2.52 | 0.31 | 1.52 | 0.57 | 19.39 | 2.43 | 38.63 | 5.67 | 85.85 | 12.69 | | |
| 54.45 | 0.52 | | | 6.54 | 4.63 | 2.57 | 0.29 | | | 23.97 | 2.90 | 46.19 | 7.14 | 93.23 | 14.69 | | |
| 52.35 | 0.59 | 0.32 | 0.19 | 8.42 | 4.94 | 2.74 | 0.32 | 1.40 | 0.64 | 34.21 | 4.00 | 37.49 | 6.14 | 71.31 | 11.83 | | |
| 50.18 | 0.66 | | | 9.95 | 3.47 | 2.77 | 0.33 | 1.86 | 0.63 | 32.15 | 3.71 | 46.62 | 6.74 | 80.14 | 12.32 | | |
| 47.94 | 0.73 | 0.50 | 0.22 | 8.98 | 4.99 | 3.14 | 0.39 | | | 39.91 | 4.60 | 56.19 | 7.74 | 106.30 | 14.97 | | |
| 45.60 | 0.81 | 0.37 | 0.10 | 8.58 | 4.25 | 3.38 | 0.39 | | | 47.22 | 5.43 | 58.86 | 6.88 | 106.35 | 15.21 | | |
| 43.16 | 0.89 | 0.63 | 0.18 | 11.36 | 4.53 | 3.68 | 0.42 | 3.10 | 1.06 | 52.18 | 6.00 | 60.01 | 8.25 | 135.35 | 18.24 | | |
| 41.43 | 0.95 | | | 8.06 | 3.42 | 3.97 | 0.46 | 3.88 | 1.10 | 0.00 | 0.00 | 69.37 | 9.34 | 144.86 | 19.00 | | |
| 40.50 | 0.98 | 0.58 | 0.17 | 9.75 | 2.53 | 4.21 | 0.50 | | | 51.63 | 5.92 | 72.56 | 9.39 | 170.97 | 22.20 | | |
| 39.55 | 1.01 | | | 9.78 | 4.48 | 4.59 | 0.52 | 2.37 | 1.03 | 48.98 | 5.61 | 53.06 | 7.67 | 170.63 | 22.03 | | |
| 38.58 | 1.04 | 0.44 | 0.16 | 9.09 | 4.31 | 4.63 | 0.54 | 2.49 | 0.74 | 45.85 | 5.31 | 61.36 | 7.85 | 175.54 | 21.35 | | |
| 37.59 | 1.07 | | | 9.30 | 2.33 | 5.44 | 0.62 | 3.31 | 1.10 | 37.59 | 4.41 | 58.65 | 7.98 | 183.16 | 22.75 | | |
| 36.57 | 1.11 | 0.53 | 0.13 | 9.76 | 1.56 | 5.50 | 0.63 | 3.54 | 0.98 | 32.20 | 3.92 | 63.31 | 8.41 | 193.88 | 24.10 | 1.72 | 0.30 |
| 35.54 | 1.14 | 0.77 | 0.18 | 9.11 | 4.11 | 6.10 | 0.70 | 0.00 | 0.00 | 23.35 | 2.89 | 55.94 | 7.52 | 190.72 | 23.31 | 1.72 | 0.30 |
| 34.47 | 1.18 | 0.75 | 0.14 | 12.15 | 2.41 | 6.93 | 0.79 | 5.64 | 1.18 | 36.69 | 4.47 | 76.10 | 9.54 | 201.78 | 24.43 | 1.72 | 0.30 |
| 33.39 | 1.21 | 0.93 | 0.16 | 11.31 | 2.12 | 8.62 | 0.98 | 5.33 | 1.04 | 27.14 | 3.26 | 88.72 | 10.40 | 209.32 | 24.57 | 1.72 | 0.30 |
| 32.27 | 1.25 | 0.74 | 0.15 | 11.61 | 3.17 | 9.55 | 1.09 | 4.53 | 1.13 | 27.86 | 3.48 | 105.66 | 12.16 | 183.14 | 22.24 | 1.72 | 0.30 |
| 31.11 | 1.29 | 0.88 | 0.14 | 10.70 | 1.76 | 13.42 | 1.52 | 5.84 | 0.90 | 37.33 | 4.39 | 139.34 | 15.80 | 189.07 | 22.00 | 1.72 | 0.30 |
| 29.93 | 1.33 | 0.85 | 0.13 | 10.61 | 1.28 | 17.31 | 1.95 | 6.51 | 0.79 | 46.73 | 5.31 | 166.56 | 18.74 | 168.52 | 19.02 | 1.72 | 0.30 |
| 28.70 | 1.37 | | | 12.37 | 1.40 | 22.15 | 2.49 | | | 55.94 | 6.53 | 189.73 | 21.31 | 175.89 | 19.77 | 1.72 | 0.30 |

**Table 5**. Experimental cross-sections for the $^{nat}Nd(p,x)$ $^{149,147,141,140,139m,138}Nd$, $^{142,138m}Pr$ and $^{139g}Ce$ reactions

| Bombarding energy | | $^{149}$Nd | | $^{147}$Nd | | $^{141}$Nd | | $^{140}$Nd | | $^{139m}$Nd | | $^{138}$Nd | | $^{142}$Pr | | $^{138m}$Pr | | $^{139}$Ce | |
|---|---|---|---|---|---|---|---|---|---|---|---|---|---|---|---|---|---|---|---|
| E (MeV) | ΔE (MeV) | σ (mb) | Δσ (mb) | σ (mb) | Δσ (mb) | σ (mb) | Δσ (mb) | σ (mb) | Δσ (mb) | σ (mb) | Δσ (mb) | σ (mb) | Δσ (mb) | σ (mb) | Δσ (mb) | σ (mb) | Δσ (mb) | σ (mb) | Δσ (mb) |
| **Series 1** | | | | | | | | | | | | | | | | | | | |
| 14.32 | 0.20 | 0.81 | 0.09 | | | | | | | | | | | 0.26 | 0.07 | | | 0.33 | 0.04 |
| 13.24 | 0.24 | 0.38 | 0.04 | | | | | | | | | | | | | | | 0.26 | 0.04 |
| 12.03 | 0.29 | 0.12 | 0.01 | | | | | | | | | | | | | | | 0.15 | 0.03 |
| 10.88 | 0.34 | 0.017 | 0.004 | | | | | | | | | | | 0.22 | 0.07 | | | 0.02 | 0.01 |
| 9.45 | 0.39 | | | | | | | | | | | | | 0.34 | 0.08 | | | 0.05 | 0.01 |
| **Series 2** | | | | | | | | | | | | | | | | | | | |
| 64.04 | 0.20 | 6.43 | 1.04 | 14.85 | 1.67 | 280.71 | 53.86 | 215.87 | 44.01 | 47.69 | 5.42 | 161.72 | 19.81 | 6.25 | 1.64 | 7.48 | 0.89 | 231.84 | 26.03 |
| 62.22 | 0.26 | | | 15.79 | 1.78 | | | 192.89 | 45.77 | | | | | | | | | 233.63 | 26.23 |
| 60.36 | 0.32 | 6.43 | 0.97 | 15.26 | 1.72 | 323.17 | 54.42 | 168.82 | 46.81 | 45.66 | 5.18 | 131.85 | 16.30 | | | 6.52 | 0.77 | 230.76 | 25.91 |
| 58.44 | 0.38 | | | 16.19 | 1.83 | | | | | | | | | 4.37 | 1.42 | | | 237.75 | 26.70 |
| 56.47 | 0.45 | 6.18 | 0.90 | 14.93 | 1.68 | 276.93 | 44.17 | 225.39 | 41.37 | 40.47 | 4.59 | 82.58 | 10.43 | 7.93 | 1.86 | 6.11 | 0.72 | 231.10 | 25.96 |
| 54.45 | 0.52 | 5.86 | 0.79 | 15.64 | 1.76 | 308.14 | 42.61 | 192.03 | 51.38 | 38.18 | 4.34 | 52.81 | 7.93 | | | 5.25 | 0.66 | 251.23 | 28.20 |
| 52.35 | 0.59 | 6.20 | 0.88 | 15.44 | 1.74 | 333.29 | 45.34 | 298.85 | 54.78 | 35.50 | 4.03 | 23.19 | 4.47 | | | 5.51 | 0.72 | 256.05 | 28.75 |
| 50.18 | 0.66 | | | 15.39 | 1.73 | | | 232.60 | 44.07 | | | | | 3.76 | 1.07 | | | 231.46 | 25.99 |
| 47.94 | 0.73 | 6.80 | 0.86 | 15.59 | 1.76 | 352.13 | 44.36 | 193.03 | 47.04 | 22.64 | 2.58 | | | | | 4.78 | 0.54 | 205.13 | 23.04 |
| 45.60 | 0.81 | 6.75 | 0.91 | 15.05 | 1.70 | 400.73 | 50.14 | | | 15.80 | 1.82 | | | | | 4.23 | 0.50 | 172.24 | 19.34 |
| 43.16 | 0.89 | 6.71 | 0.99 | 14.63 | 1.65 | 379.29 | 45.98 | 227.51 | 47.44 | 8.72 | 1.02 | 3.25 | 2.22 | | | 3.25 | 0.42 | 104.07 | 11.68 |
| 41.43 | 0.95 | | | 13.60 | 1.54 | | | 227.54 | 45.05 | 0.00 | 0.00 | | | 5.72 | 1.50 | 0.00 | 0.00 | 70.22 | 7.89 |
| 40.50 | 0.98 | 5.78 | 1.08 | 13.63 | 1.54 | 365.64 | 45.04 | 234.91 | 52.24 | 3.50 | 0.42 | 0.88 | 1.67 | | | 2.63 | 0.37 | 47.38 | 5.35 |
| 39.55 | 1.01 | | | 13.13 | 1.48 | | | 227.91 | 46.42 | | | | | | | | | 29.33 | 3.30 |
| 38.58 | 1.04 | 6.26 | 0.95 | 12.72 | 1.43 | 355.75 | 43.69 | 263.27 | 48.26 | 1.21 | 0.21 | 4.38 | 2.28 | 5.03 | 1.64 | 2.78 | 0.38 | 16.35 | 1.87 |
| 37.59 | 1.07 | | | 12.72 | 1.44 | | | 224.87 | 53.78 | | | | | 5.62 | 1.32 | | | 12.51 | 1.43 |
| 36.57 | 1.11 | 6.49 | 0.77 | 11.10 | 1.26 | 347.31 | 41.68 | | | | | 2.38 | 1.55 | | | 2.18 | 0.28 | 8.31 | 0.95 |
| 35.54 | 1.14 | 6.70 | 1.11 | 10.85 | 1.23 | 340.79 | 42.17 | | | | | | | | | 2.10 | 0.35 | 6.29 | 0.75 |
| 34.47 | 1.18 | 7.04 | 0.92 | 10.46 | 1.20 | 313.78 | 38.26 | 182.49 | 41.91 | | | | | 3.91 | 1.11 | 1.81 | 0.25 | 5.21 | 0.63 |
| 33.39 | 1.21 | 6.90 | 0.81 | 10.26 | 1.17 | 364.13 | 43.42 | 161.77 | 37.59 | | | | | | | 1.53 | 0.21 | 5.80 | 0.69 |
| 32.27 | 1.25 | 6.44 | 0.89 | 9.33 | 1.08 | 307.09 | 38.85 | 108.02 | 27.60 | | | | | | | 1.24 | 0.21 | 3.97 | 0.51 |
| 31.11 | 1.29 | 6.59 | 0.81 | 9.65 | 1.09 | 328.29 | 39.99 | 65.61 | 16.66 | | | | | | | 1.12 | 0.16 | 4.75 | 0.58 |
| 29.93 | 1.33 | 6.38 | 0.75 | 8.93 | 1.00 | 313.85 | 38.61 | 22.80 | 4.30 | | | | | | | 0.77 | 0.13 | 3.30 | 0.41 |
| 28.70 | 1.37 | | | 8.78 | 0.99 | | | | | | | | | | | | | 3.04 | 0.36 |

### *3.1.1 Production cross sections of $^{150}$Pm*

The experimental data for production of $^{150}$Pm ($T_{1/2}$=2.68 h) are shown in Fig. 1. The agreement with the experimental data in the literature is good in the overlapping energy range. The theoretical prediction in TENDL-2015 does not represent well the measured values.

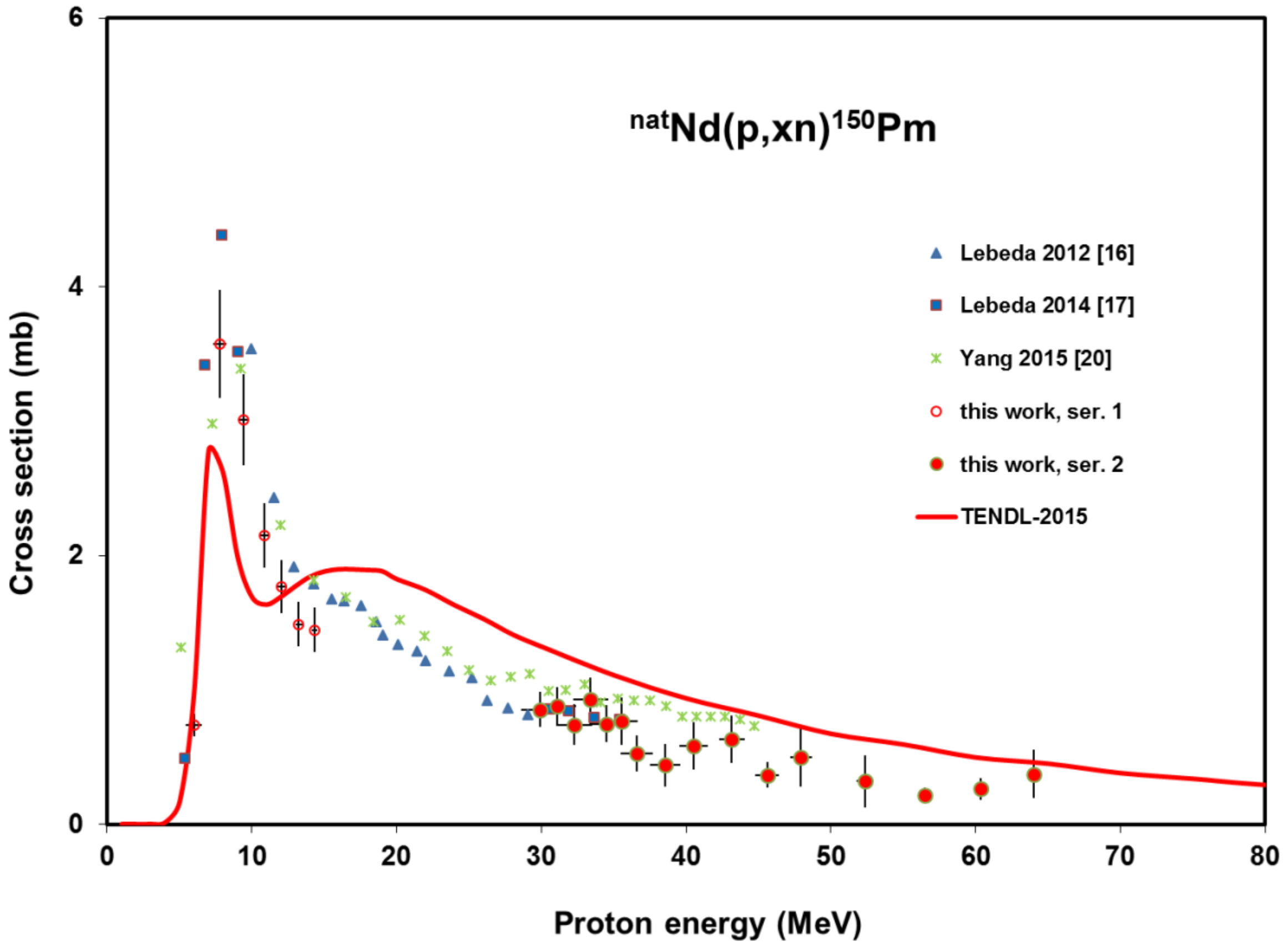

Fig.1 Experimental and theoretical cross sections for the $^{nat}$Nd(p,x)$^{150}$Pm reaction

### *3.1.2 Production cross sections of $^{149}$Pm*

The radionuclide $^{149}$Pm ($T_{1/2}$ = 53.08 h) is produced directly and through the $^{149}$Pr ($T_{1/2}$=2.26 min) → $^{149}$Nd ($T_{1/2}$ = 1.728 h) → $^{149}$Pm decay chain. Our measured experimental $^{149}$Pm cross-sections are cumulative, obtained after the complete decay of $^{149}$Nd. The agreement with the cumulative cross- section data of Lebeda et al. [16,17], Yang et al. [19] and with the TENDL-2015 prediction is acceptable good (Fig. 2)  Only the $^{150}$Nd(p,2n) reaction contributes to the production.

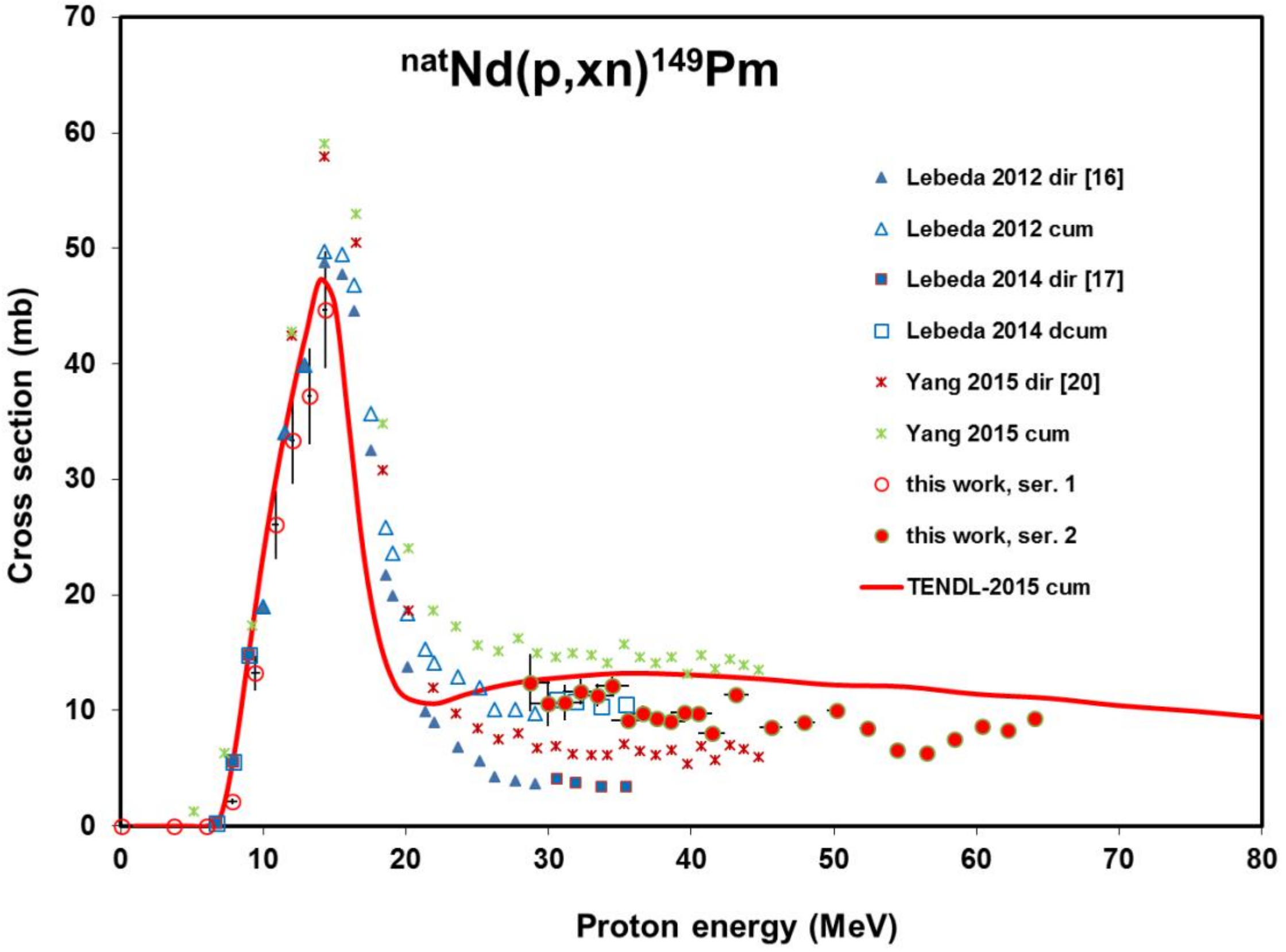

Fig.2 Experimental and theoretical cross sections for the $^{nat}$Nd(p,x)$^{149}$Pm  reaction

### *3.1.3 Production cross sections of $^{148m}$Pm*

The $^{148}$Pm has two long-lived isomeric states: the $^{148m}$Pm ($T_{1/2}$ = 41.29 d) and the $^{148g}$Pm ($T_{1/2}$ = 5.368 d). The higher laying state has only small internal decay fraction (IT: 4.2 %). There is good agreement with the earlier experimental data (Fig. 3). The $^{148}$Nd(p,n) and the $^{150}$Nd(p,3n) reactions contribute to the production of both isomeric states. The TENDL predictions systematically underestimate the $^{150}$Nd(p,3n) formation of the metastable state (see also discussion for ground state).

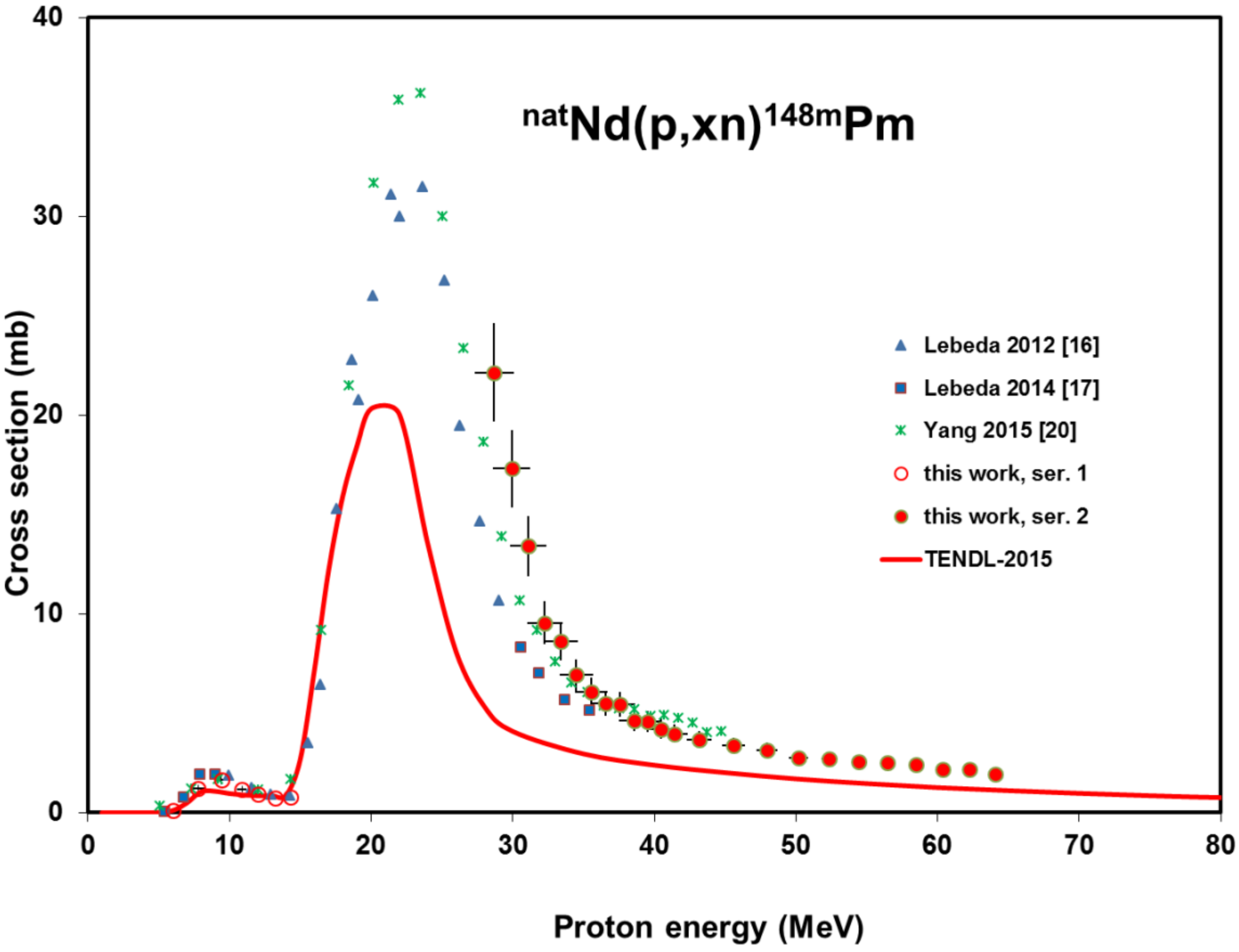

Fig.3 Experimental and theoretical cross sections for the $^{nat}$Nd(p,x)$^{148m}$Pm reaction

### *3.1.4 Production cross sections of $^{148g}$Pm*

The cross sections for the direct production of the $^{148g}$Pm ($T_{1/2}$ = 5.368 d) (after correction for the small contribution from the decay of the $^{148m}$Pm) are shown in Fig. 4 in good agreement with literature values. TENDL-2015 significantly overestimates the $^{150}$Nd(p,3n) process (not good representation of distribution of reaction cross-section over ground- and meta-state).

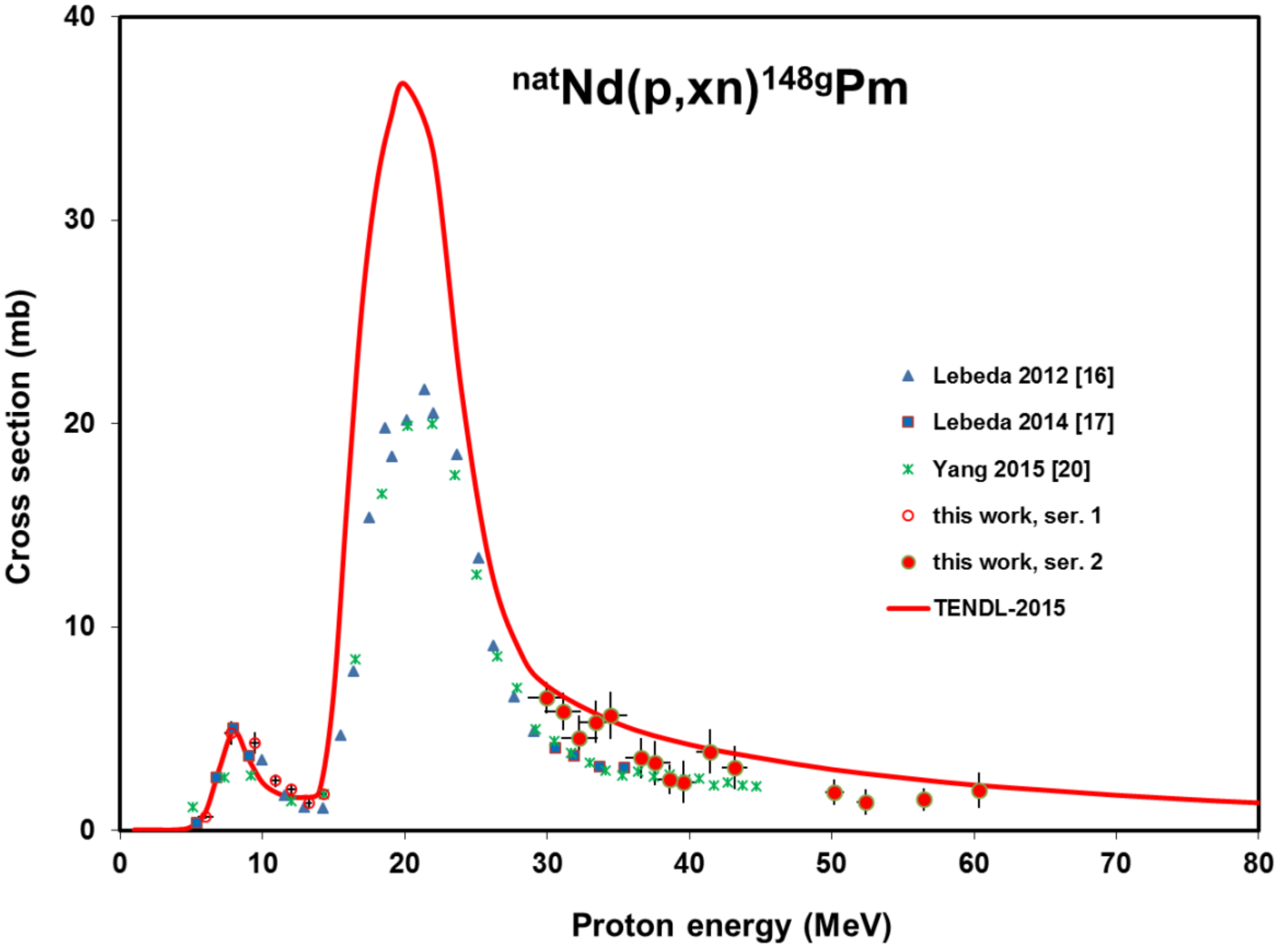

Fig.4 Experimental and theoretical cross sections for the $^{nat}$Nd(p,x)$^{148g}$Pm reaction

### *3.1.5 Production cross sections of $^{146}$Pm*

Three reactions participate in the production of $^{146}$Pm ($T_{1/2}$ = 5.53 y) by proton irradiation of $^{nat}$Nd as shown in the excitation function with three distinct maxima. There is an energy shift between the earlier experimental data set of Lebeda [16,17], but the agreement is considered acceptable by comparing to other activation measurements of Yang [19] (Fig. 5). TENDL-2015 shows an energy shift becoming more obvious with increasing bombarding energy.

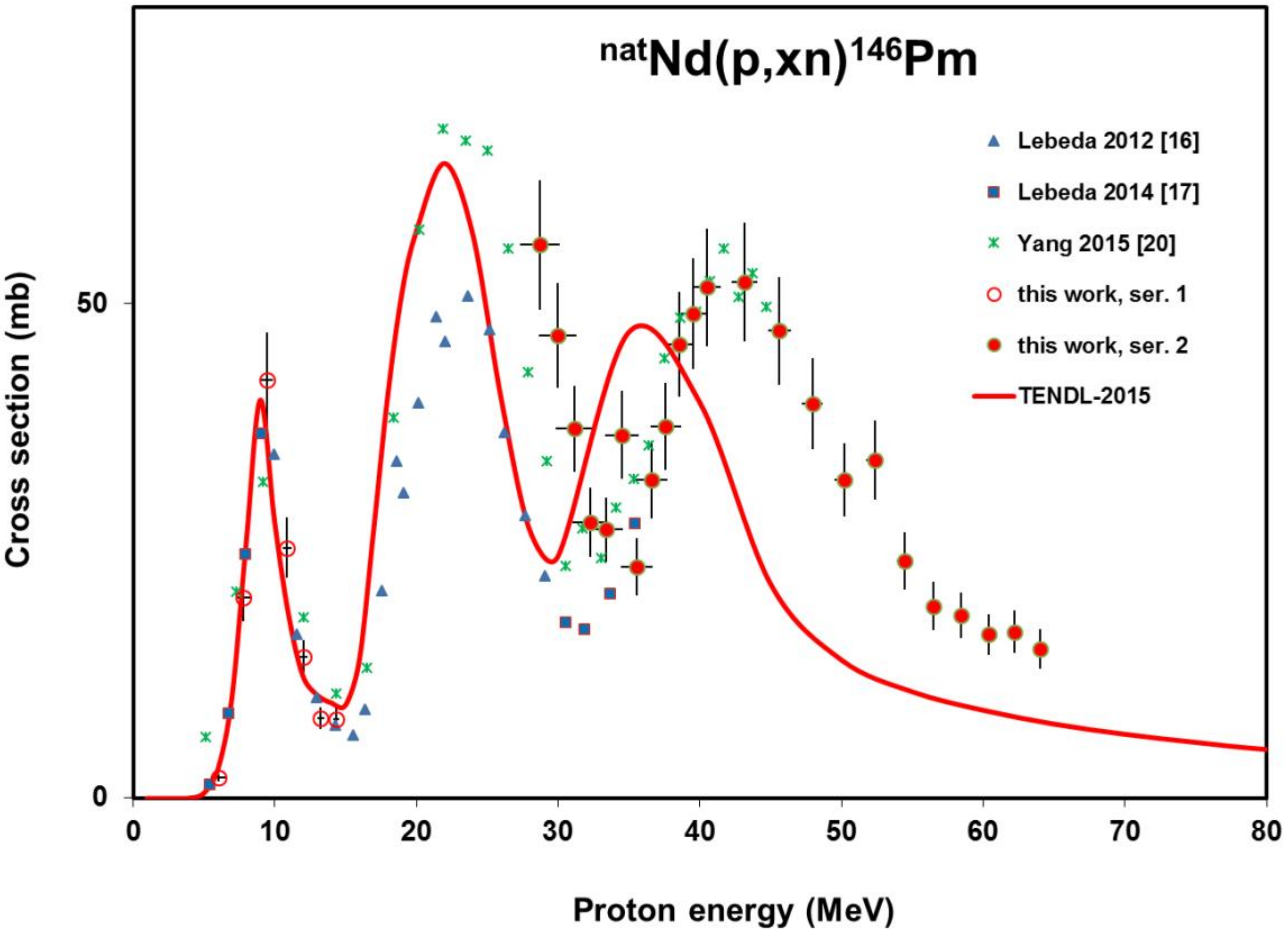

Fig.5 Experimental and theoretical cross sections for the $^{nat}$Nd(p,x)$^{146}$Pm reaction

### *3.1.6 Production cross sections of $^{144}$Pm*

The maximum of first four contributing reactions of $^{144}$Pm ($T_{1/2}$= 365 d) can be seen in the Fig. 6. The agreement between all experimental datasets of Yang [19] and also with the theory is good both in shape and in magnitude.

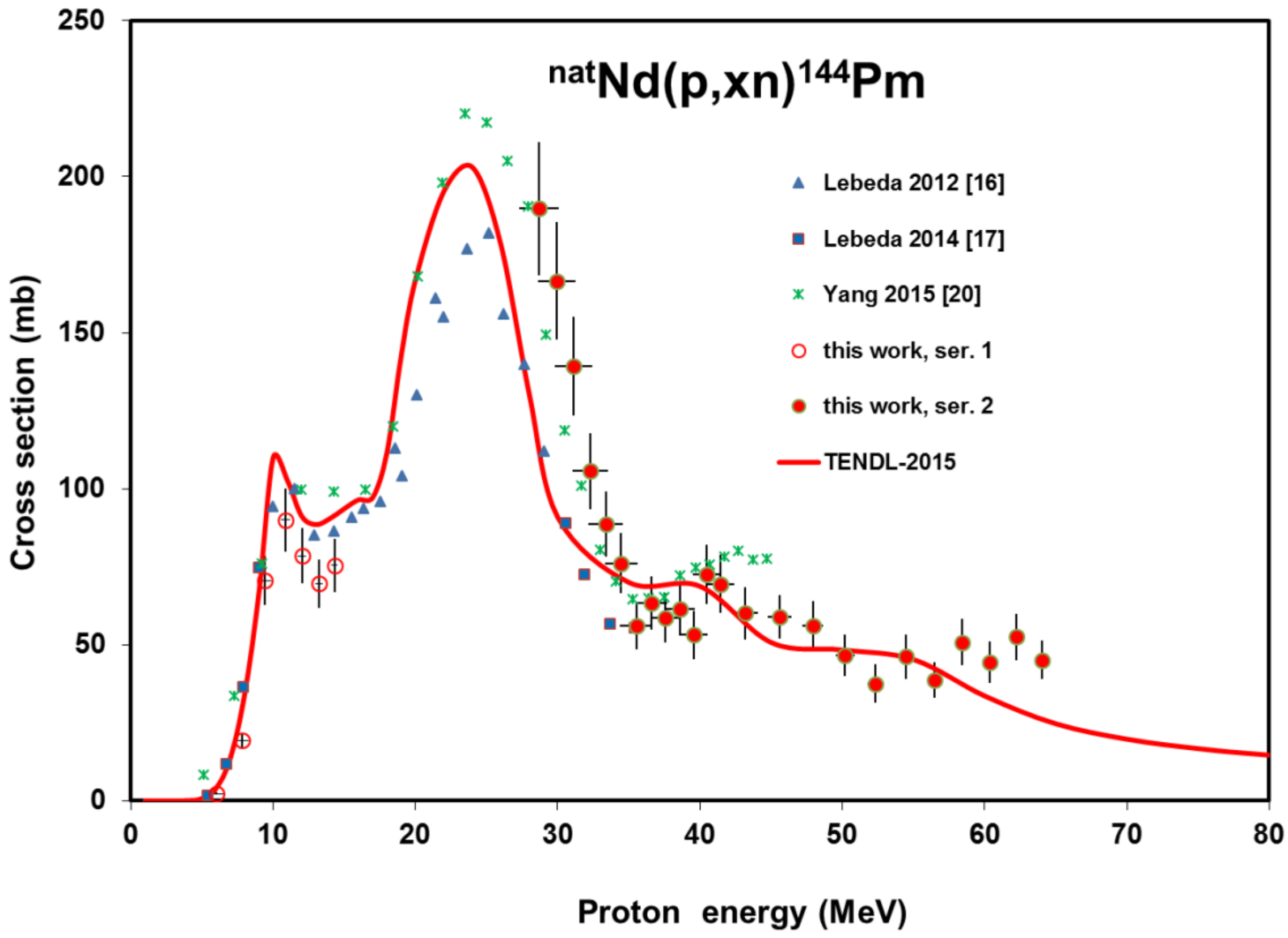

Fig.6 Experimental and theoretical cross sections for the $^{nat}$Nd(p,x)$^{144}$Pm reaction

### *3.1.7 Production cross sections of $^{143}$Pm*

The (p,xn) reactions on stable isotopes of $^{143,144,145,146,148,150}$Nd participate in the production of $^{143}$Pm ($T_{1/2}$ = 265 d). Our results are close to results of Lebeda et al. [16,17]. The data of Yang et al. [20] are systematically higher by around 30 % (Fig. 7). The description by TENDL-2015 is acceptable.

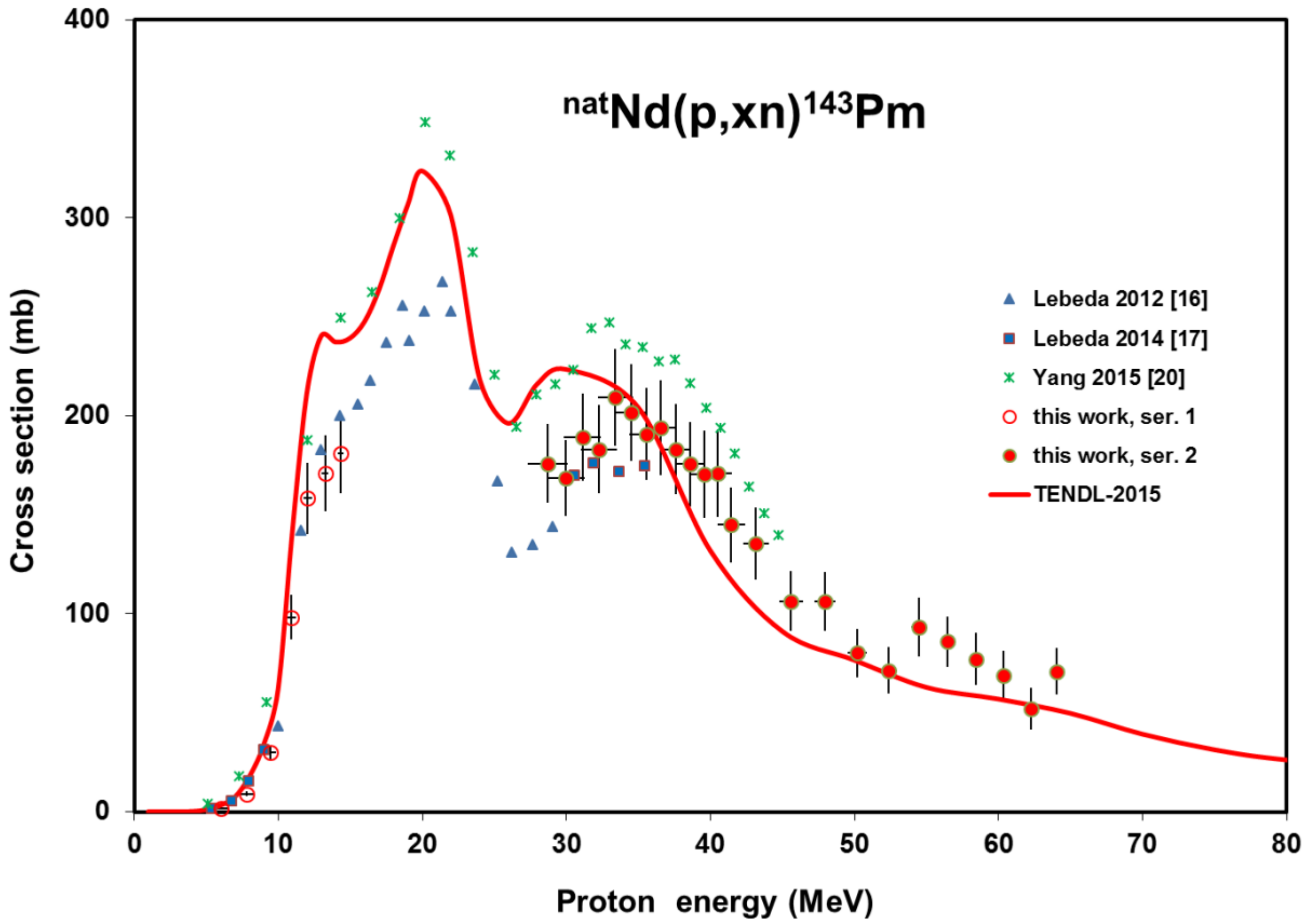

Fig.7 Experimental and theoretical cross sections for the $^{nat}$Nd(p,x)$^{143}$Pm reaction

#### *3.1.8 Production cross sections of $^{141}$Pm*

Due to the relatively short half-life ($T_{1/2}$ = 20.9 min) under our experimental circumstances we could get only two cross section data points in the low energy irradiation near the reaction threshold (Fig. 8).

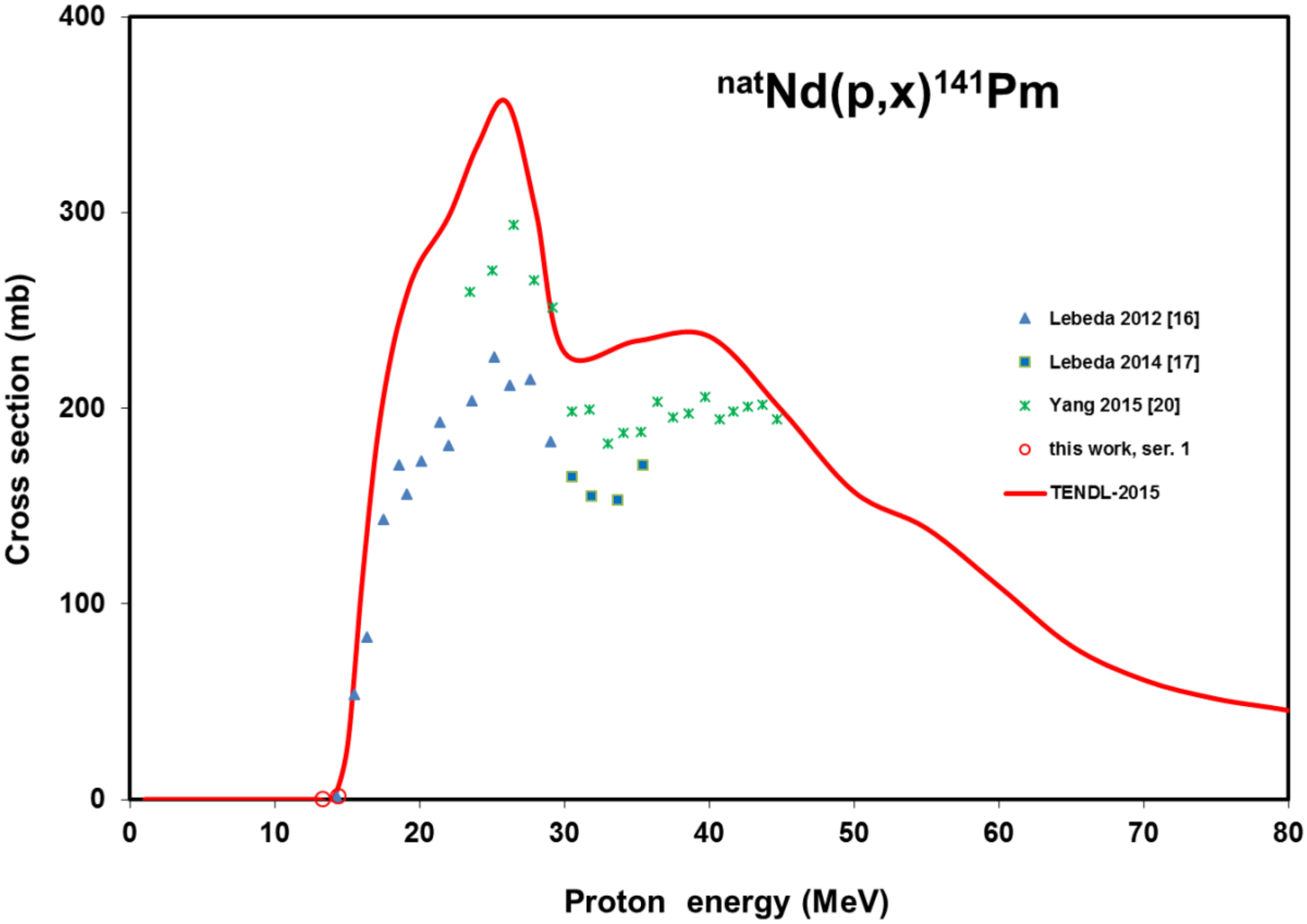

Fig.8 Experimental and theoretical cross sections for the $^{nat}$Nd(p,x)$^{141}$Pm reaction

### 3.1.9 *Production cross sections of $^{149}Nd$*

The measured cross sections refer to cumulative production of $^{149}$Nd ($T_{1/2}$ = 1.728 h): directly via $^{150}$Nd(p,pn) reaction and from the complete decay of the short-lived $^{149}$Pr ($T_{1/2}$ = 2.26 min). According to the theory the contribution from the $^{149}$Pr decay is small. The tendency and the magnitude of the experimental and theoretical data are similar, but there is systematic shift in the energy (Fig. 9). The shape of the TENDL curve is acceptable but it underestimates the experiment under 25 MeV and overestimates above this energy.

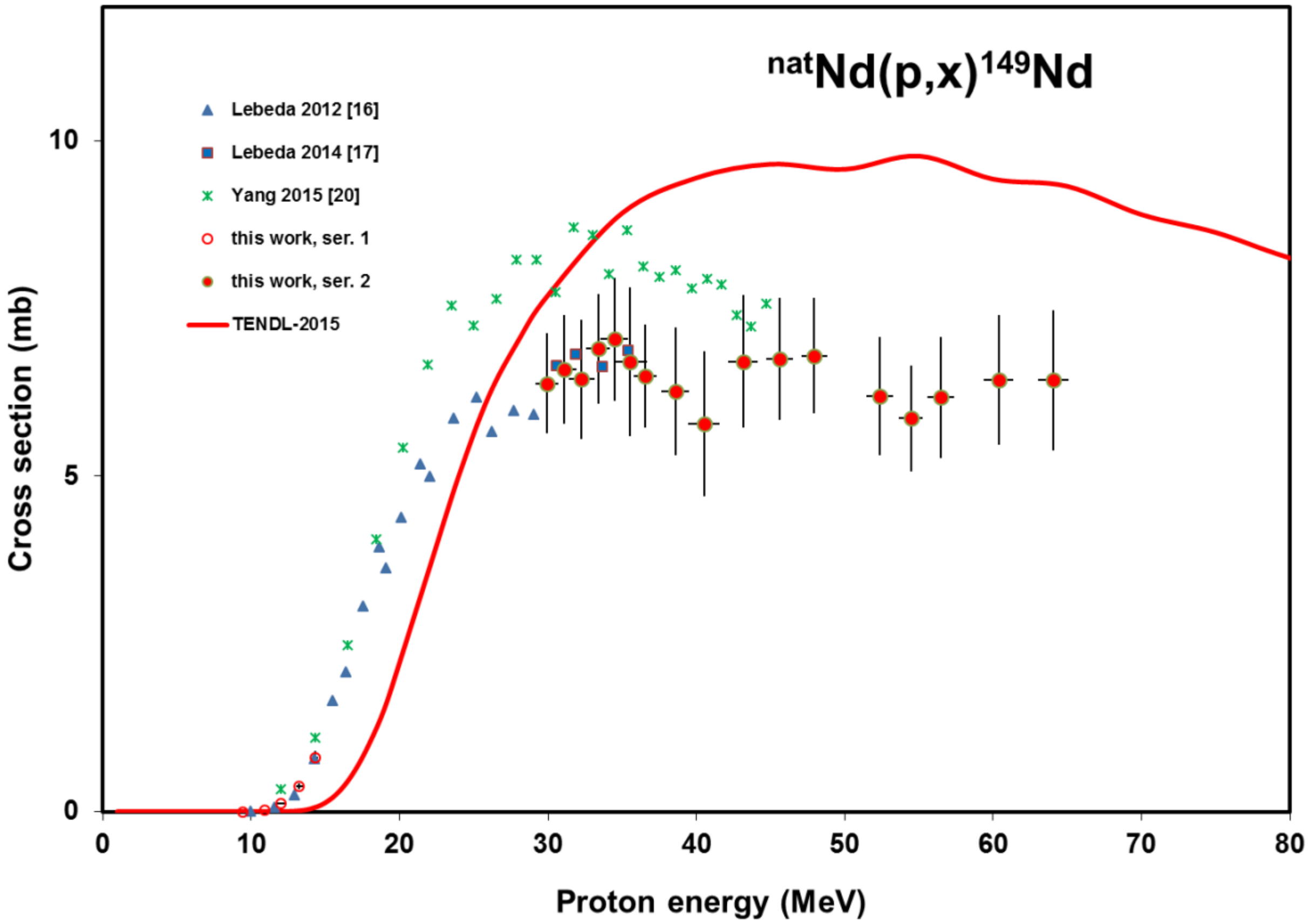

Fig.9 Experimental and theoretical cross sections for the $^{nat}$Nd(p,x) $^{149}Nd$ reaction

*3.1.10 Production cross sections of $^{147}Nd$*

For production of $^{147}$Nd ($T_{1/2}$ = 10.98 d) the earlier and the present results show excellent agreement (Fig. 10). The cross sections are cumulative, as a contribution from short-lived $^{147}$Pr ($T_{1/2}$ = 13.4 min) exists. At higher energies, above 30 MeV, the theory overestimates the experimental data.

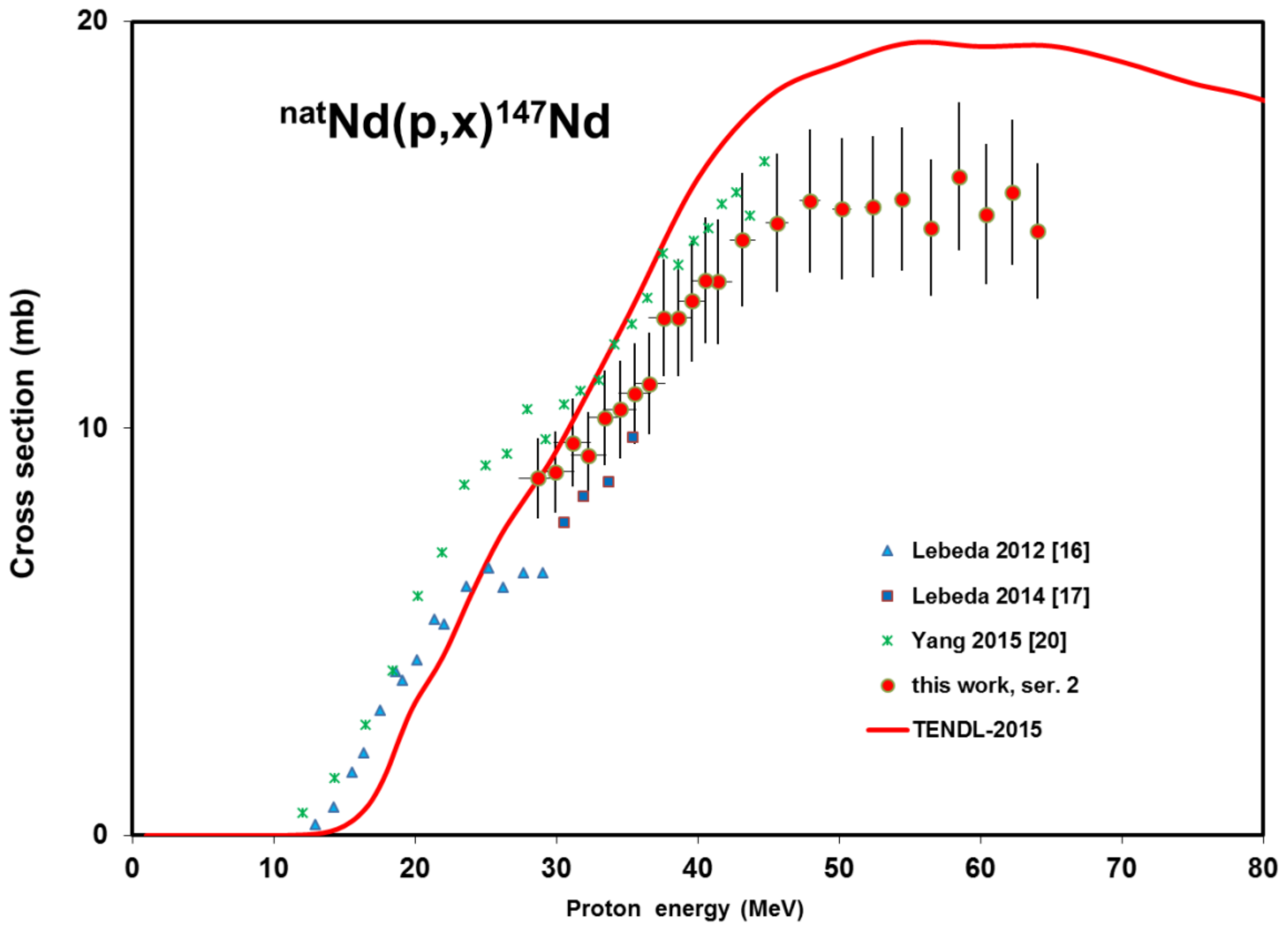

Fig.10 Experimental and theoretical cross sections for the $^{nat}$Nd(p,x) $^{147}Nd$ reaction

### *3.1.11 Production cross sections of $^{141}$Nd*

The ground state of $^{141}$Nd ($T_{1/2}$ = 2.49 h) is produced directly, from $^{141}$Pm (EC decay, $T_{1/2}$ = 20.90 min) and through the short-lived $^{141m}$Nd ($T_{1/2}$= 62.0 s, IT: 99.95 %) isomeric state. We have deduced cumulative cross section data (Fig 11). Lebeda et al. [16,17] reported independent cross section data for production of $^{141}$Nd (m+).To get cumulative data for comparison we summarized the $^{141}$Pm and $^{141}$Nd cross-section data published in [16,17] (Fig. 11), and got acceptable agreement. The approximation of the TENDL theoretical data is acceptable good.

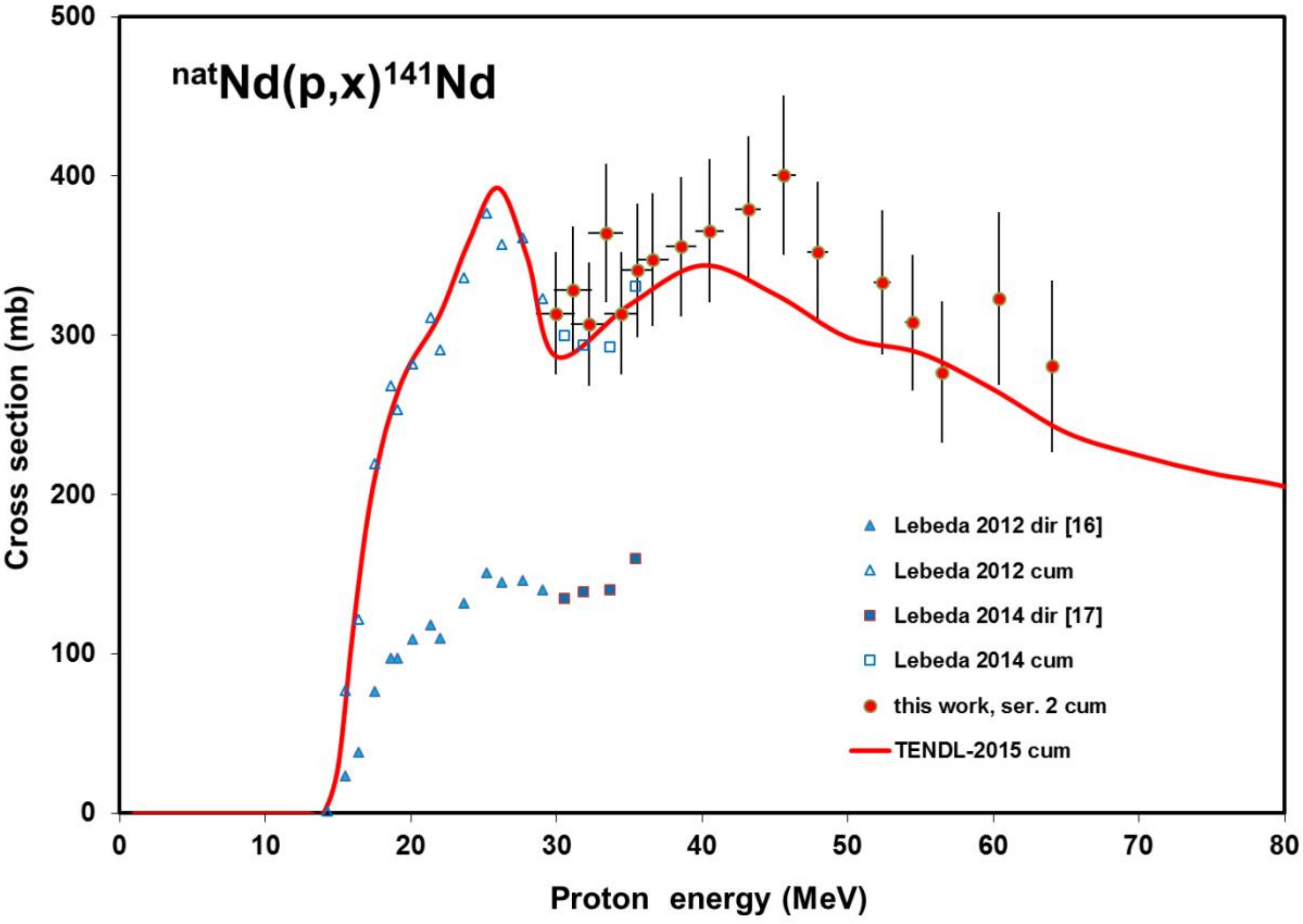

Fig.11 Experimental and theoretical cross sections for the $^{nat}$Nd(p,x) $^{141}$*Nd* reaction

### *3.1.12 Production cross sections of $^{140}Nd$*

The $^{140}Nd$ ($T_{1/2}$ = 3.37 d, ε: 100 %) is produced directly and through the decay of $^{140}Pm$ ($T_{1/2}$ = 9 s, ε: 100 %). As $^{140}Nd$ has no gamma-lines, the cross-sections were determined through assessment of the short half-life $^{140}Pr$ ($T_{1/2}$ = 3.39 min) daughter isotope. In the comparison with earlier data of Lebeda et al. [16,17], a slight energy shift can be observed near the threshold (Fig. 12). An even larger shift is seen in the TENDL-2015 prediction in the opposite direction.

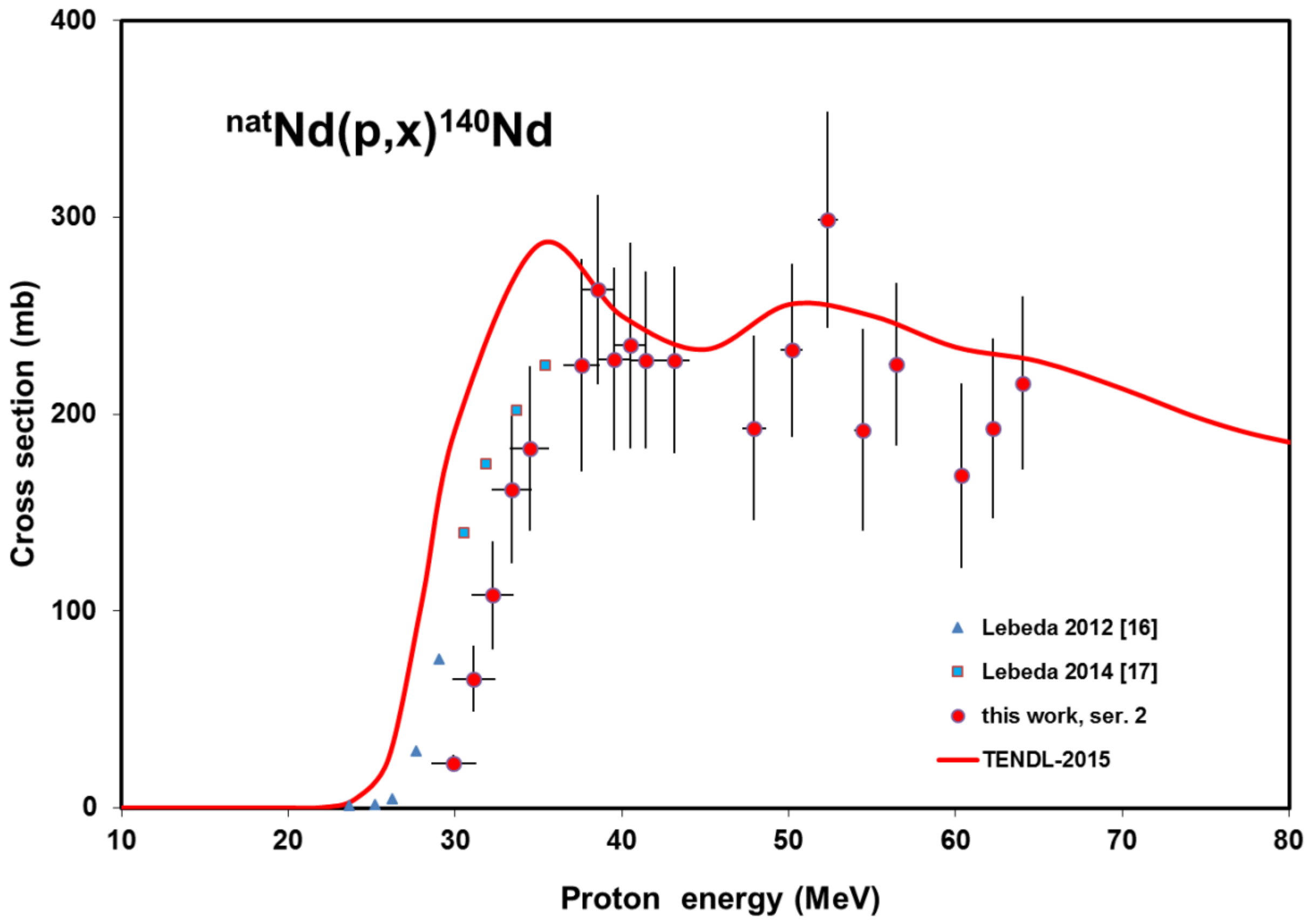

Fig.12 Experimental and theoretical cross sections for the $^{nat}Nd(p,x)$ $^{140}Nd$ reaction

### *3.1.13 Production cross sections of $^{139m}Nd$*

Out of the two long-lived isomeric states we obtained cross sections for production of $^{139m}Nd$ higher lying high spin state ($T_{1/2}$ = 5.5 h). This state is produced only directly as $^{139}Pm$ ($T_{1/2}$ = 4.15 min) and decays only to the ground state $^{139g}$Nd ($T_{1/2}$ = 29.7 min). The comparison with the theory (energy shift) and the data of Yang [20] (good agreement) is shown in Fig. 13.

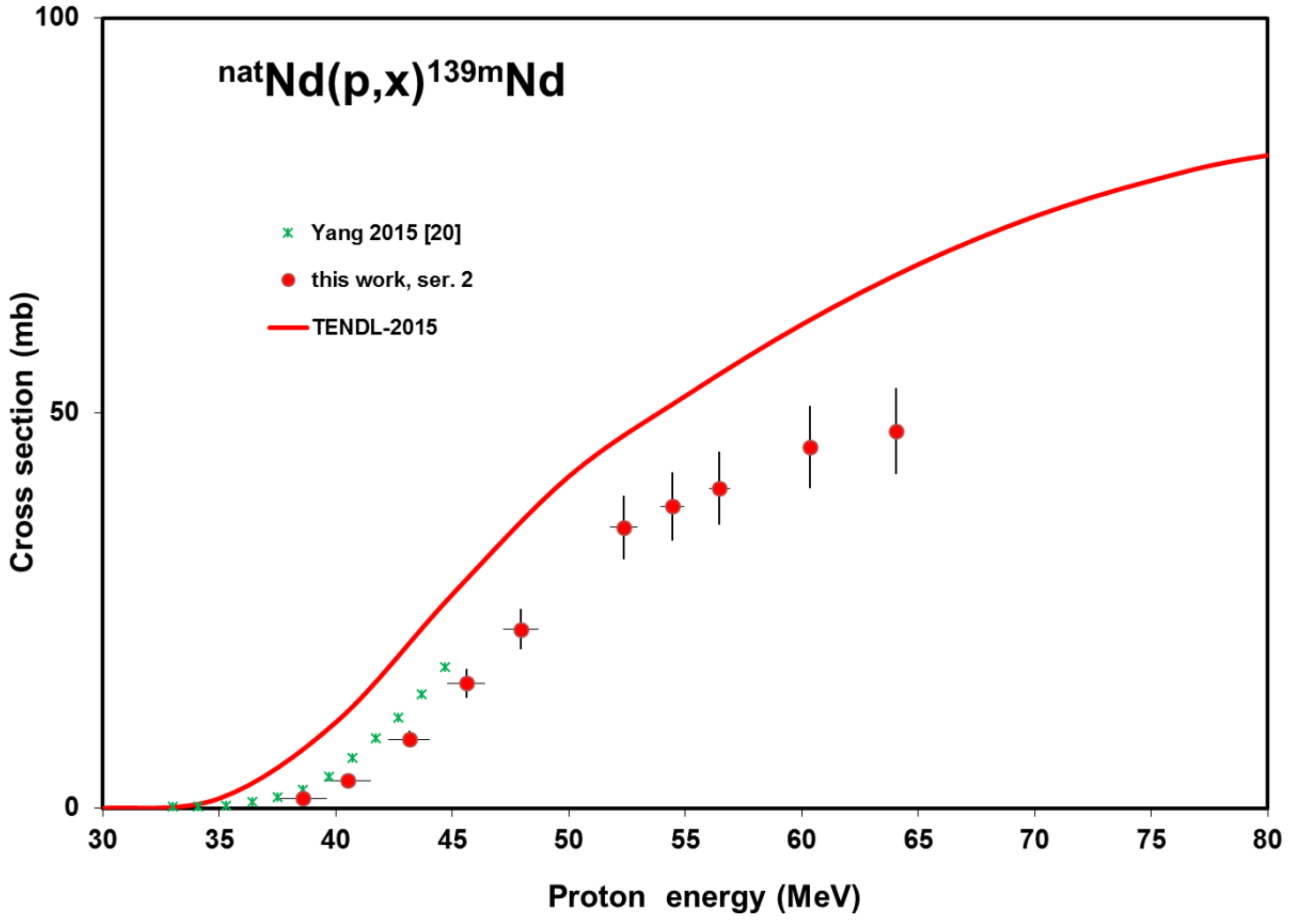

Fig.13 Experimental and theoretical cross sections for the $^{nat}$Nd(p,x)$^{139m}$Nd reaction

### *3.1.14 Production cross sections of $^{138}$Nd*

No earlier experimental data were found for the production cross-sections of $^{138}$Nd ($T_{1/2}$ = 5.04 h). The measured cross-sections are cumulative they include the contribution from the $^{138}$Pm decay ($T_{1/2}$ = 3.24 min, ε: 100%) (Fig.14). Energy shift with TENDL-2015 predications can be observed.

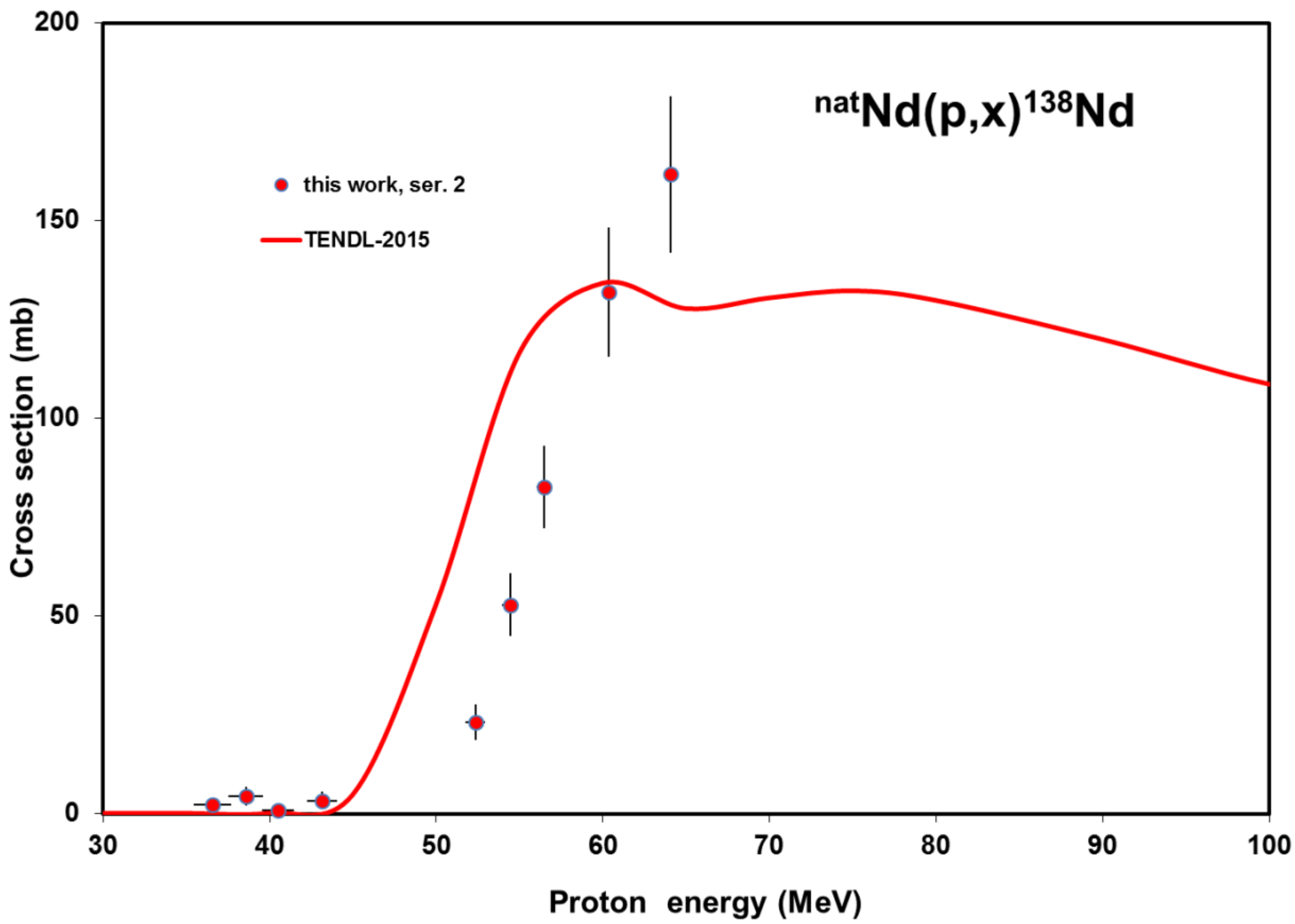

Fig.14 Experimental and theoretical cross sections for the $^{nat}$Nd(p,x)$^{138}$Nd reaction

### *3.1.15 Production cross sections of $^{142}$Pr*

$^{142}$Pr is a closed shell isotope. It is produced via (d,2pxn) reactions. It has two long-lived isomeric states: the $^{142g}$Pr ($T_{1/2}$ = 19.12 h) ground state and the $^{142m}$Pr ($T_{1/2}$ =14.6 min) metastable state, which decays by IT 100% to the ground state. The measured cross sections are cumulative (m+). They include the contribution from the decay of the isomeric state. Our data are systematically higher, compared to the literature experimental data in the overlapping energy range, but the uncertainty of our results are high (Fig. 15).

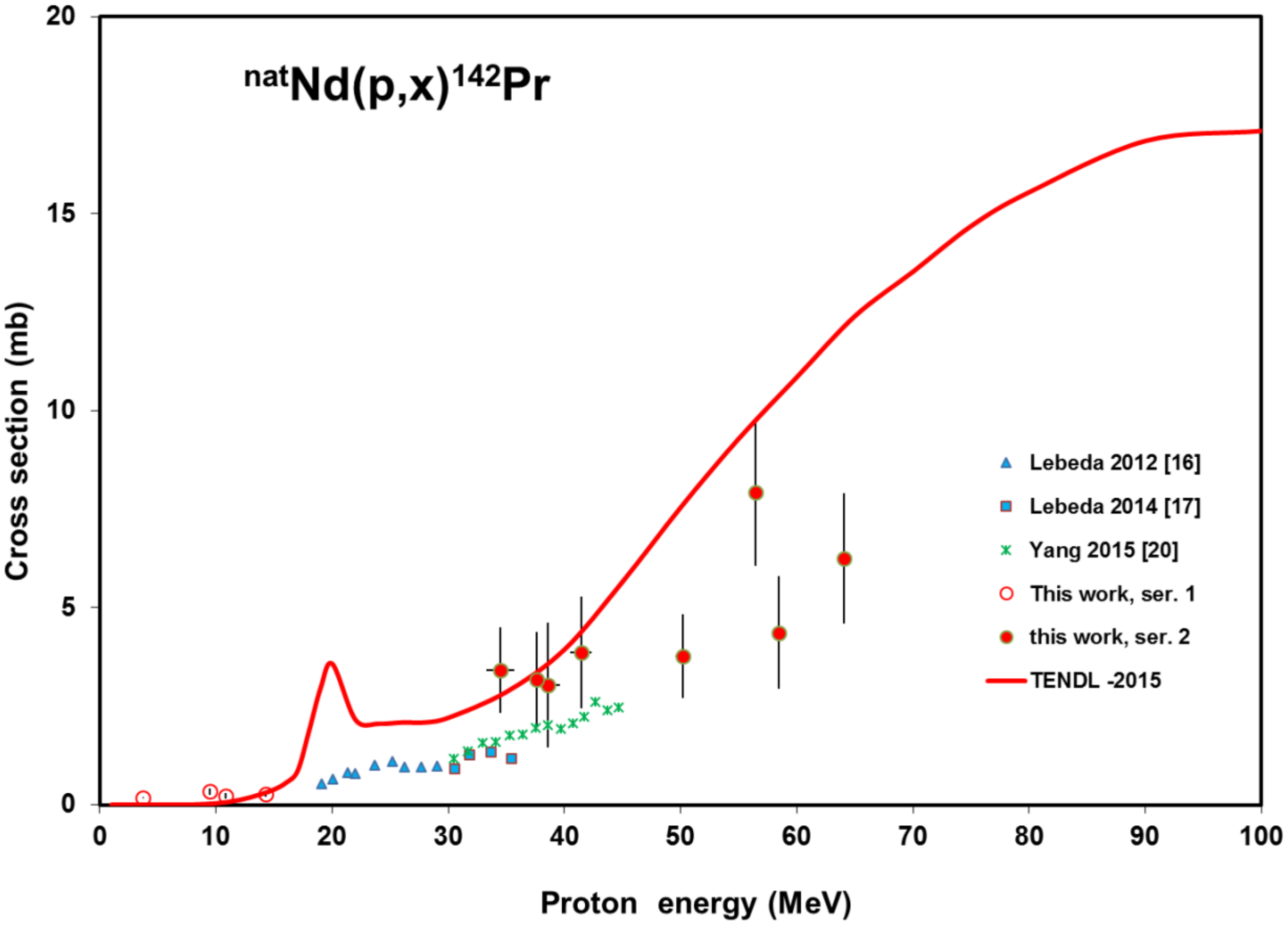

Fig.15 Experimental and theoretical cross sections for the $^{nat}$Nd(p,x)$^{142}$Nd reaction

### *3.1.16 Production cross sections of $^{138m}$Pr*

Out of the longer-lived states, we deduced independent production cross-sections for $^{138m}$Pr ($T_{1/2}$ = 2.12 h) as isobaric possible parent $^{138}$Nd ($T_{1/2}$ = 5.04 h) decays only to $^{138g}$Pr ($T_{1/2}$ = 1.45 min). The agreement with the earlier experimental data and with the theory is acceptable (Fig. 16).

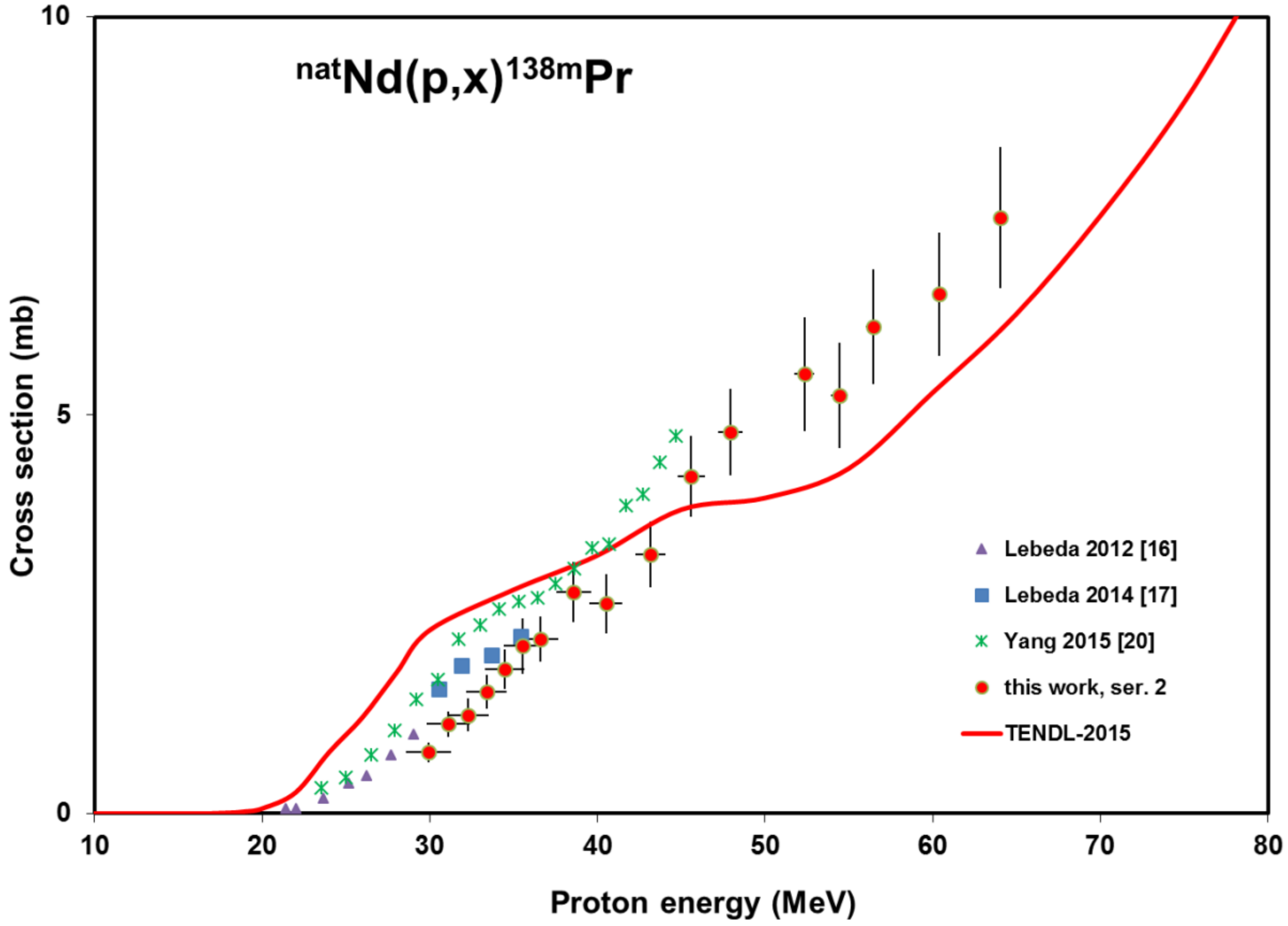

Fig.16 Experimental and theoretical cross sections for the $^{nat}$Nd(p,x)$^{138m}$Pr reaction

### *3.1.17 Production cross sections of $^{139g}$Ce*

The main contributions of the measured cumulative cross-sections for the long-lived ground state $^{139g}$Ce ($T_{1/2}$=137.641 d) arise from the shorter-lived $^{139}$Pm ($T_{1/2}$ = 4.15 min) → $^{139}$Nd ($T_{1/2}$ = 5.5h; 29.7 min) → $^{139}$Pr ($T_{1/2}$ = 4.41 h) parent decay chain (Fig. 17). The direct production cross-sections through the $^{142}$Nd(p,p3n) reaction are very small. The theory follows the trend of the experimental curves, the agreement with the earlier experimental results is acceptable.

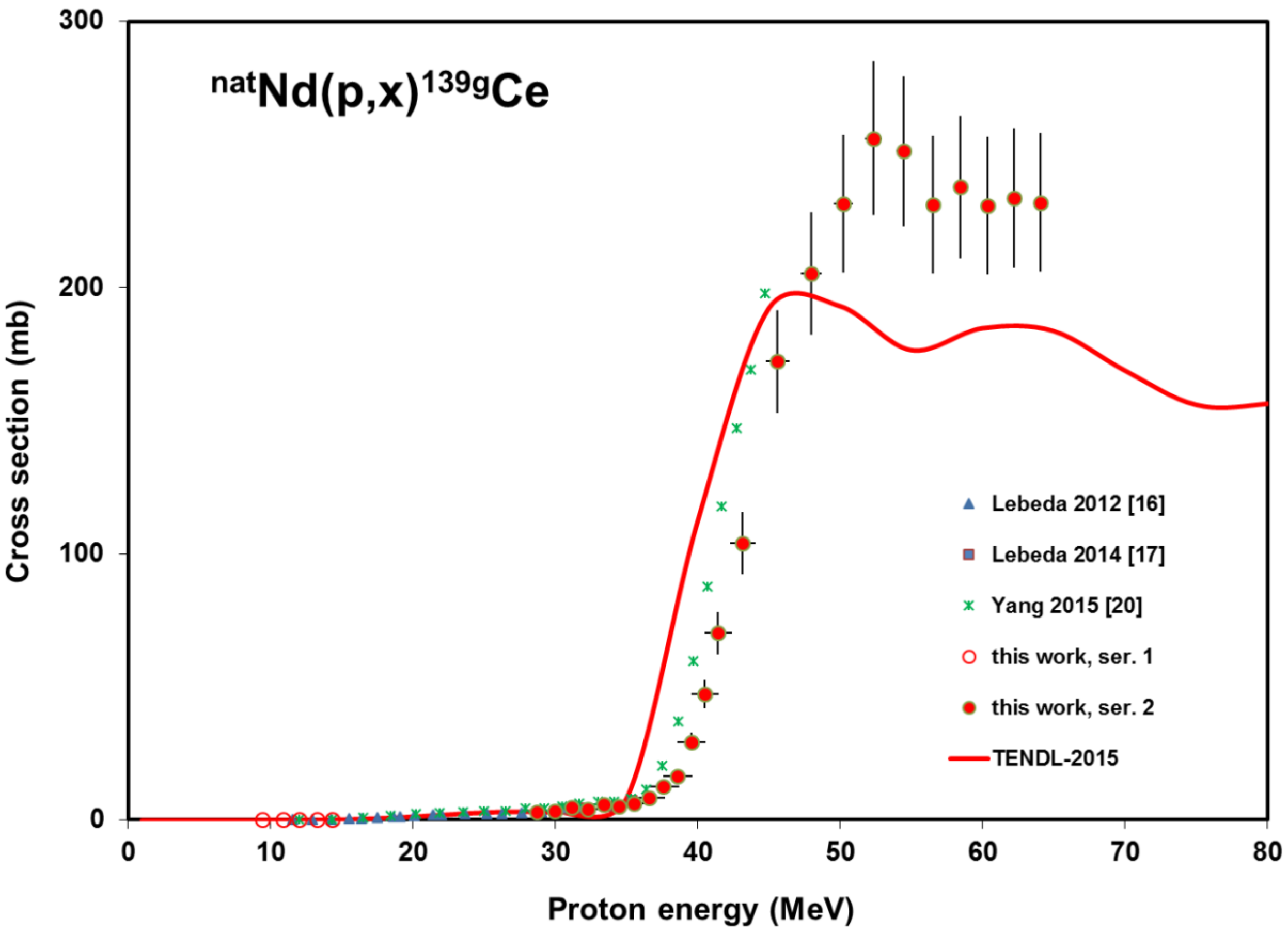

Fig.17 Experimental and theoretical cross sections for the $^{nat}$Nd(p,x) $^{139g}$Ce reaction

## 3.2 INTEGRAL YIELDS

The calculated integral yields (integrated yield for a given incident energy down to the reaction threshold) are shown in Figs. 18-19 in comparison with experimental integral thick target yields found in the literature. Our thick target yields were calculated from fited curves to our experimental cross-section data and some literature data with acceptable agreement with ours in order to fill the hole between our two series. In the case of Yang data [19] thin target yields were published, which should be converted into thick target yields. The results represent so called physical yields (obtained in an instantaneous irradiation time) [32,33]. In Fig. 18. A good agreement is seen between our and the literature data, except in the lower energy region, where the literature cross section data are also higher. It is also seen that the difference in the low energy (and low cross section) region does not influence a lot the agreements in the higher energy region in most cases. In Fig. 19, in the case of Nd, Pr and Ce radioisotopes the literature data of Yang [19] are systematically higher, which is the consequence that the cross section data are also higher.

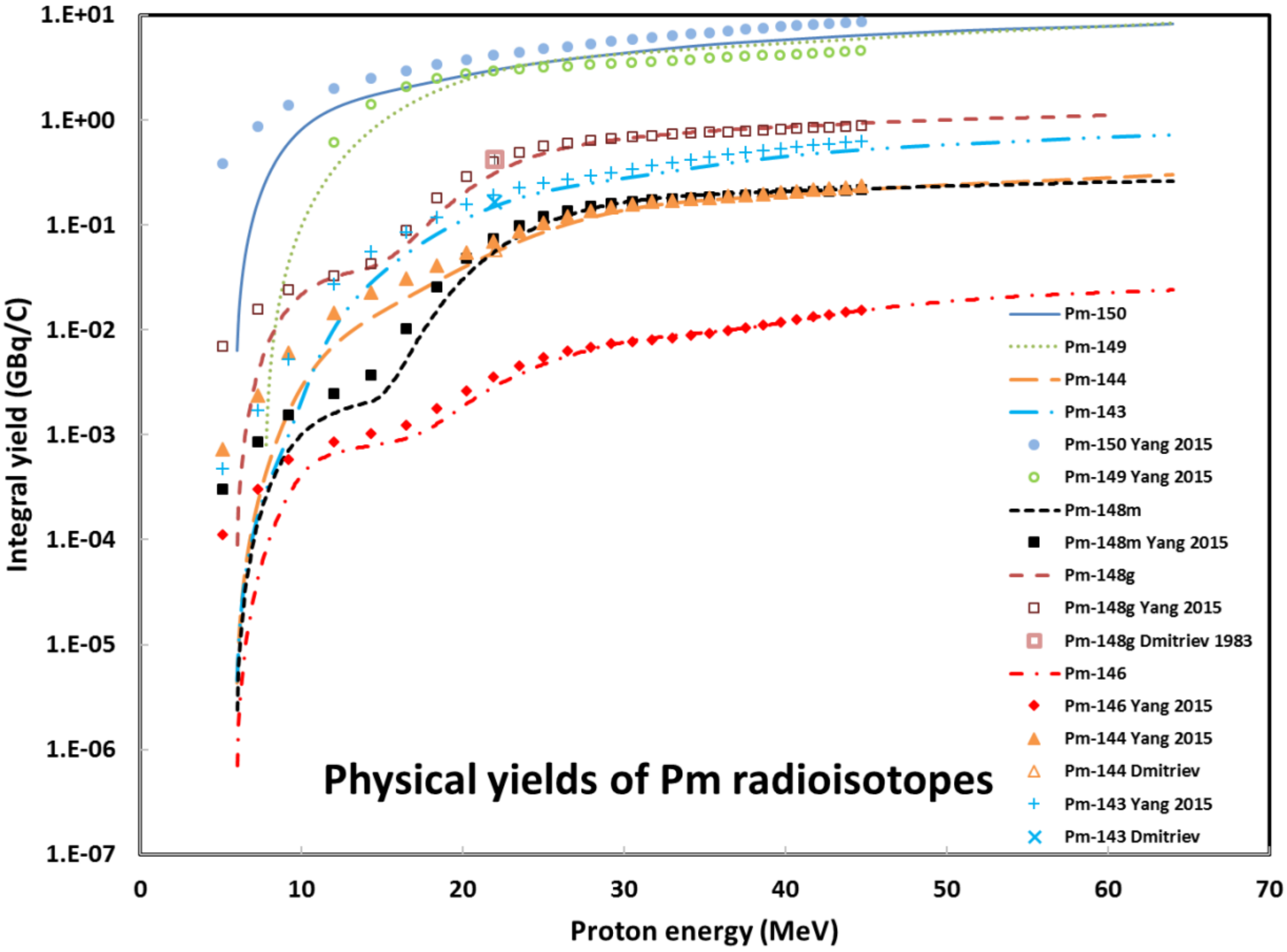

Fig. 18 Integral yields for the of ${}^{nat}Nd(p,x)\,{}^{150,149,148m,148g,146,144,143}Pm$, nuclear reactions

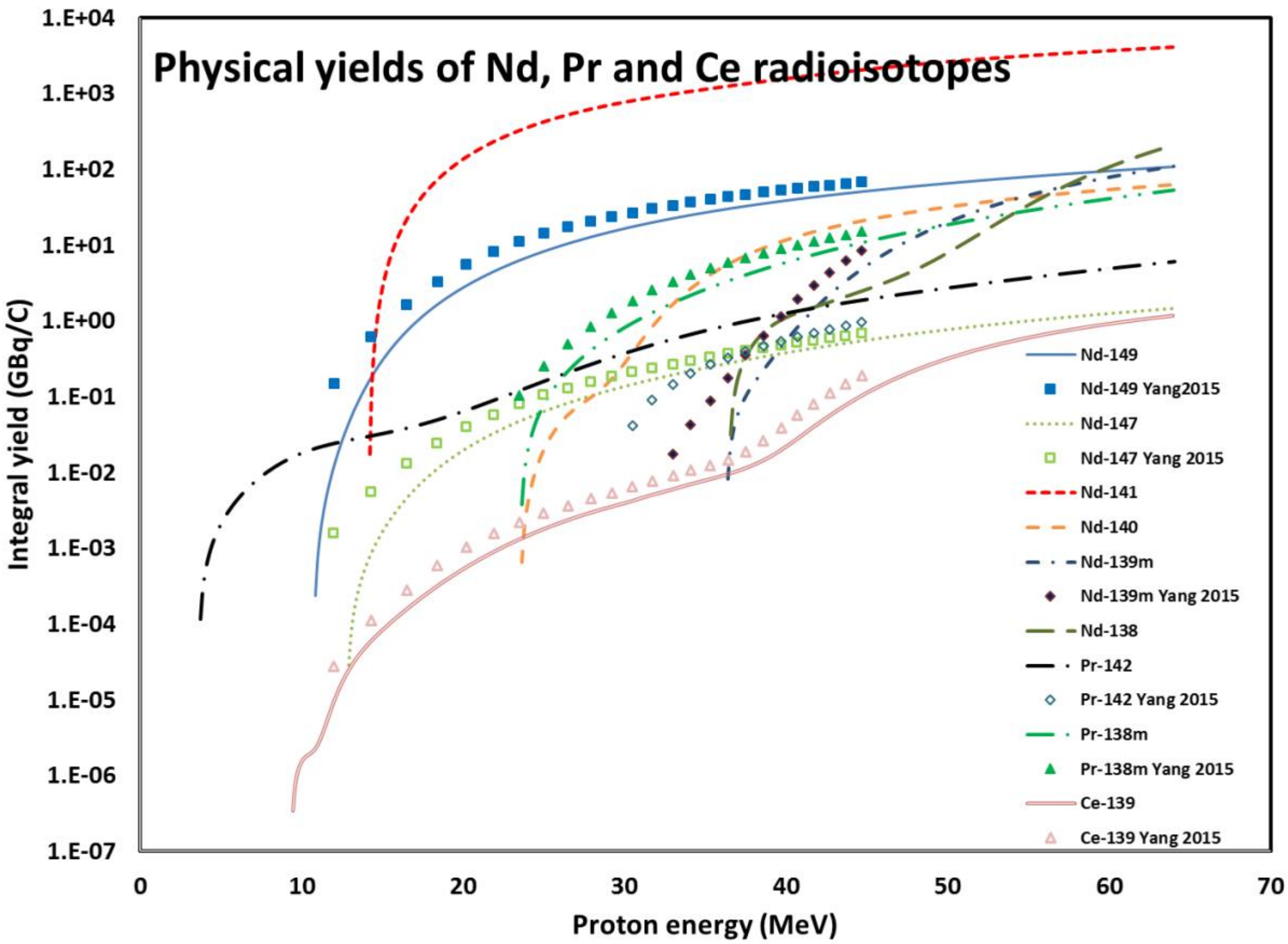

Fig. 19 Integral yields for the $^{nat}Nd(p,x)$ $^{149,147,141,140,139m,138}Nd$, $^{138m}Pr$ and $^{139g}Ce$ nuclear reactions

## 4. APPLICATIONS

Radioactive lanthanides have a great potential in nuclear medicine, both in radiotherapy and diagnostic. Out of the investigated reaction products the $^{149}$Pm ($T_{1/2}$ = 53.1 h ) and $^{140}$Nd ($T_{1/2}$ = 3.37 d) radioisotopes have potential interest in nuclear medicine while $^{139}$Ce is used as a calibration source. The measured data can be used for optimization of the production routes of these radionuclei.

### 4.1 Production of $^{149}$Pm (2.21 d)

The moderate beta energy emitter $^{149}$Pm ($\beta^{-}_{max}$ : 1.071 MeV) can be produced carrier free at reactors via the $^{148}$Nd(n,γ)$^{149}$Nd (1.73 h) →$^{149}$Pm ($T_{1/2}$ = 53 h) production route. Also the charged particle reactions $^{148}$Nd(d,n)$^{149}$Pm, $^{150}$Nd(d,3n) and $^{150}$Nd(p,2n) result in n.c.a. production. Our new data were measured above 30 MeV and below 15 MeV, which is not the optimum energy range for the (p,2n) reaction.

### 4.2 Production of $^{140}$Nd (3.37 d)

The Auger electron emitter $^{149}$Nd has good nuclear properties with potential for endo-radiotherapeutic applications. The decay of $^{140}$Pr (3.39 min) daughter offers the possibility of using positron emission tomography (PET) to quantify $^{140}$Nd. When using the presently investigated proton induced reactions on Nd the yield would be high, but the product is carrier added.

### 4.3 Production of $^{139}$Ce (137.641 d)

$^{139}$Ce is a long-lived radioisotope emitting a single high intensity, low energy, γ-line, optimal for applications connected to calibration of different detectors used in nuclear medicine and nuclear physics.

It can be be produced directly and indirectly via various reactions: $^{139}$La(p,n), $^{nat}$La(p,x), $^{nat}$La(d,x) $^{nat}$Ce(p,x)$^{139}$Pr, $^{139}$Ce, $^{nat}$Ce(d,x)$^{139}$Pr→$^{139}$Ce, $^{139}$Ce(d,2n), $^{141}$Pr(p,x), $^{141}$Pr(d,x), $^{nat}$Nd(d,x) and $^{nat}$Nd(p,x). According to Fig. 19 the $^{nat}$Nd(p,x) reaction has a high production yield at high energy accelerators (100-40 MeV).

## 5. Summary and conclusions

In this work proton induced cross sections were measured on natural neodymium targets up to 75 MeV bombarding energy in two series. Because of a failure in the irradiation plan the values between 14 and 29 MeV are missing from our measurements. All cross section data reported here are new above 45 MeV. Our new data were compared with the literature and in some cases the agreement was acceptable good but also considerable disagreement and even energy shift were seen in several cases. The results of nuclear reaction model calculations made by the TALYS code and taken from the TENDL-2015 on-line library show a good prediction in most cases as far as the shape of the excitation functions regarded, but fail in the magnitude of the curves and also a considerable energy shift is seen in several cases. There are such isotopes (e.g. $^{141}$Nd), where both the shape and the values of the TENDL-2015 results show acceptable agreement with our and with the literature data.

From the measured excitation functions thick target physical yields were also deduced and compared with the literature data. In the case of Pm radioisotopes generally an acceptable agreement was seen between our data and the literature results, except the low energy region, while in the case of Nd, Pr and Ce radioisotopes an energy/magnitude shift can be observed in the cases where the literature data are available.

Applications of some candidate isotopes for nuclear medicine ($^{149}$Pm, $^{140}$Nd, $^{139}$Ce) were also discussed emphasizing on the possible production routes.

## Acknowledgements

This work was done in the frame of MTA-FWO (Vlaanderen) research projects. The authors acknowledge the support of research projects and of their respective institutions in providing the materials and the facilities for this work.

**Figure captions**

Fig.1 Experimental and theoretical cross sections for the $^{nat}Nd(p,x)^{150}Pm$ reaction

Fig.2 Experimental and theoretical cross sections for the $^{nat}Nd(p,x)^{149}Pm$ reaction

Fig.3 Experimental and theoretical cross sections for the $^{nat}Nd(p,x)^{148m}Pm$ reaction

Fig.4 Experimental and theoretical cross sections for the $^{nat}Nd(p,x)^{148g}Pm$ reaction

Fig.5 Experimental and theoretical cross sections for the $^{nat}Nd(p,x)^{146}Pm$ reaction

Fig.6 Experimental and theoretical cross sections for the $^{nat}Nd(p,x)^{144}Pm$ reaction

Fig.7 Experimental and theoretical cross sections for the $^{nat}Nd(p,x)^{143}Pm$ reaction

Fig.8 Experimental and theoretical cross sections for the $^{nat}Nd(p,x)^{141}Pm$ reaction

Fig.9 Experimental and theoretical cross sections for the $^{nat}Nd(p,x)$ $^{149}Nd$ reaction

Fig.10 Experimental and theoretical cross sections for the $^{nat}Nd(p,x)$ $^{147}Nd$ reaction

Fig.11 Experimental and theoretical cross sections for the $^{nat}Nd(p,x)$ $^{141}Nd$ reaction

Fig.12 Experimental and theoretical cross sections for the $^{nat}Nd(p,x)$ $^{140}Nd$ reaction

Fig.13 Experimental and theoretical cross sections for the $^{nat}Nd(p,x)$ $^{139m}Nd$ reaction

Fig.14 Experimental and theoretical cross sections for the $^{nat}Nd(p,x)^{138}Nd$ reaction

Fig.15 Experimental and theoretical cross sections for the $^{nat}Nd(p,x)^{142}Nd$ reaction

Fig.16 Experimental and theoretical cross sections for the $^{nat}Nd(p,x)^{138m}Pr$ reaction

Fig.17 Experimental and theoretical cross sections for the $^{nat}Nd(p,x)$ $^{139g}Ce$ reaction

Fig. 18 Integral yields for the of $^{nat}Nd(p,x)$ $^{150,149,148m,148g,146,144,143}Pm$, nuclear reactions

Fig. 19 Integral yields for the $^{nat}Nd(p,x)$ $^{149,147,141,140,139m,138}Nd$, $^{138m}Pr$ and $^{139g}Ce$ nuclear reactions